\documentclass[conference]{IEEEtran}
\IEEEoverridecommandlockouts
\usepackage{cite}
\usepackage{amsmath,amssymb,amsfonts}
\usepackage{algorithmic}
\usepackage{graphicx}
\usepackage{textcomp}
\usepackage{xcolor}
\usepackage{dblfloatfix}
\usepackage{caption}
\usepackage{subcaption}
\def\BibTeX{{\rm B\kern-.05em{\sc i\kern-.025em b}\kern-.08em
    T\kern-.1667em\lower.7ex\hbox{E}\kern-.125emX}}
\begin{document}

\makeatletter
\newcommand{\linebreakand}{%
  \end{@IEEEauthorhalign}
  \hfill\mbox{}\par
  \mbox{}\hfill\begin{@IEEEauthorhalign}
}
\makeatother

\title{Effects of Hybrid CPU and
Cache Architectures on Parallel HPC/Cloud Applications
\thanks{This work is funded by Pacific Northwest National Laboratory as part of internship.}}

\author{

\IEEEauthorblockN{
Nanda Velugoti \IEEEauthorrefmark{1} \and
Joesph Manzano \IEEEauthorrefmark{2} \and
Nathan Tallent \IEEEauthorrefmark{2} \and
Kyle Hale \IEEEauthorrefmark{1}
} \\
\linebreakand
\IEEEauthorblockA{\IEEEauthorrefmark{1} 
\textit{Department of Computer Science} \\
\textit{Illinois Institute of Technology}\\
Chicago, USA \\
}
\and
\IEEEauthorblockA{\IEEEauthorrefmark{2}
\textit{High Performance Computing Group} \\
\textit{Pacific Northwest National Laboratory}\\
Richland, Washington \\
}
}

\maketitle

\thispagestyle{plain}
\pagestyle{plain}

\begin{abstract}
Hybrid CPU architectures have entered the mainstream desktop computing with the announcement of Intel's Alderlake architecture. Such a transition to heterogeneous CPU architecture has various performance and power implications on existing parallel workloads. In this paper we study the effects and impact of hybrid core and cache architecture on the performance of highly parallel HPC workloads. We also illustrate interesting thread scaling behaviour for parallel workloads and describes the reason for such behaviour both qualitatively and quantitatively. We also explore the impact of hybrid cache architecture on parallel shared-data HPC applications. Finally, we illustrate that 1) parallel applications with work imbalance (i.e., threads in application perform different amount of work) scale better across hybrid cores when thread affinity is disabled and 2) hybrid cache architecture has very little impact on parallel shared-data applications except for some workloads with locks. This work lays the foundation for our future work which focuses on extending this work to model parallel workloads and hybrid CPU architectures to improve their performance in terms of execution time, memory usage and power consumption.
\end{abstract}

\begin{IEEEkeywords}
HPC, hybrid, heterogeneous, architecture
\end{IEEEkeywords}

\section{Introduction}

HPC and scientific computing applications have highly parallel data-intensive, compute-intensive workloads that require a huge amount of compute, power and memory resources. Moreover these applications are traditionally deployed on supercomputers and in some rare situations they may be deployed on servers/cloud architectures. However, these architectures usually tend to be homogeneous in nature and are specifically designed to be highly parallel and closely connect clusters of compute units. This is because in the HPC applications benefit from homogeneous systems with minimal OS noise (data parallelism, frequent bulk synchronization).
Hybrid/heterogeneous CPU architectures are not new in the landscape of computing. The idea was popularized a decade ago and gave birth to the embedded and mobile computing revolution. However, with the announcement of Intel's Alderlake CPU architecture, hybrid CPU architectures have entered the mainstream desktop computing. This transition to hybrid CPU design comes with various performance and power trade-offs depending on the workload. There are other downsides of hybrid architectures. Architects may worry that they require increased architectural complexity for memory systems, inter-core networks, and possibly instruction sets. For example, Alder Lake’s E-cores have AVX 256 but not AVX512. Further, users primarily concerned with performance may avoid hybrid systems. Regardless of these concerns, we tackle the research question: how do HPC workloads scale in the hybrid CPU system? We answer this by quantitatively measuring the performance of HPC workloads across hybrid cores and qualitatively analyze the effects of hybrid cores and caches on these workloads to arrive at some exciting insights.

\begin{figure*}[!t]
    \centering
    \includegraphics[width=\textwidth,keepaspectratio]{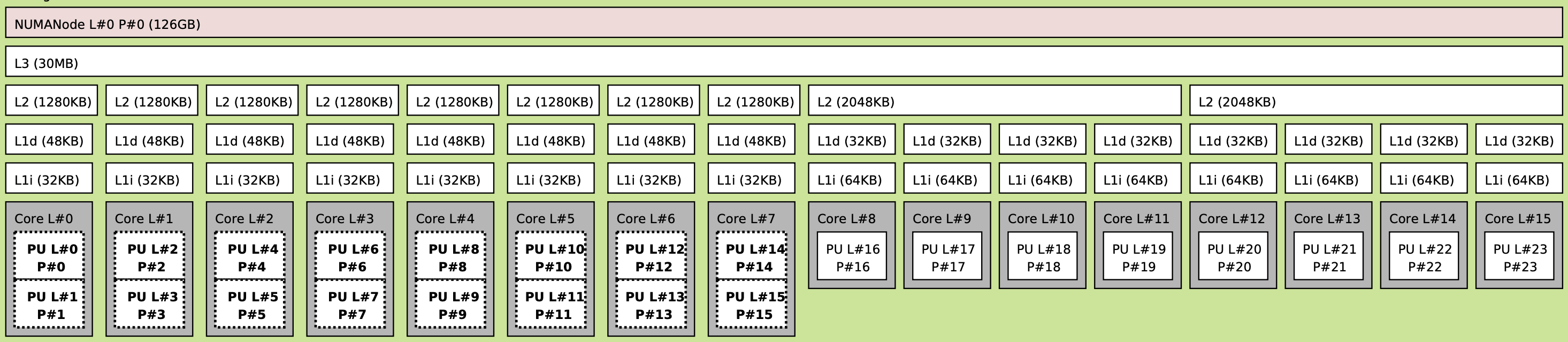}
    \caption{Output of Linux's \textit{lstopo} command for Intel's 12\textsuperscript{th} generation big.LITTLE type hybrid CPU architecture.}
    \label{fig:alderlake-arch}
\end{figure*}

\section{Background}

Before going into the details of the study of effects of hybrid CPUs, it is better to go through the hybrid CPU architecture under consideration. Fig \ref{fig:alderlake-arch} shows the output of Linux's \textit{lstopo} command, which details the topology/layout of the CPU architecture (in this case, Alderlake architecture). Intel's Alderlake CPU architectures consists of: 1) 8 P-cores (Cores L\#0-7), capable of hyper-threading; 2) 8 E-cores (Cores L\#8-15) with no hyper-threading features. Thus, in total there are 16 physical cores and 24 (8*2 + 8) logical cores. Here, we can easily notice the interesting big.LITTLE type CPU design with P/E-cores, where P-core are more capable (i.e., higher operating frequency) when compared to E-cores. P-cores max operating clock speed is 6500 MHz where as for E-cores it is 3900 MHz, thus E-cores have about $ \frac{2}{3}^{rd} $ the capacity of P-cores.

Another interesting aspect of this architecture is the heterogeneous cache layout. Again, Fig \ref{fig:alderlake-arch} shows L1 and L2 caches are private to P-cores, however, in case of E-cores only L1 caches are private and two sets of E-cores of size 4 share a common L2 cache.

\section{Motivation}
Currently, Alderlake architecture is part of their mobile/desktop class. The idea of exploring the effects of Hybrid CPU architecture on highly parallel HPC workloads is a novel one. The goal here is to study how parallel workloads scale across the hybrid cores and what. This type of study could lead to interesting insights about how one could leverage heterogeneous architecture to improve the performance of parallel workloads.
To quantitatively and qualitatively analyze the effects of hybrid CPU on parallel workloads we found two interesting aspects to focus on: 
\begin{enumerate}
    \item How does a hybrid core architecture impact the performance of parallel workloads? 
    \item What is the impact of hybrid (L2) cache architecture on parallel data-shared workloads?
\end{enumerate}
Thus we end up with two aspects of studying the hybrid architectures: 1) studying the impact of hybrid cores and 2) studying the impact of hybrid cache.

\section{Aspect 1: Impact of Hybrid Cores}
To understand the impact of hybrid cores on highly parallel workloads, we need to carefully: 1) setup a test environment;  2) choose the relevant workloads varying degrees of parallelism.

\subsection{Methodology}
The basic idea here is to understand how parallel workload scaling looks like on hybrid cores. To do this, we can scale the no. of threads given to a parallel workload and capture the performance in terms of the execution time. 

Another important detail here is that, it is common to disable hyper-threading while running parallel HPC workloads. We achieve this effect using OMP\_PLACES parameter, which is discussed further in the upcoming section. 

\subsection{Test Setup}

First, we choose workloads that are relevant in HPC and are parallel in nature. We have chosen \textit{miniVite} \cite{bib-minivite} a graph based workload that implements a community detection approach known as the Louvain method. We have also tested SP, BT, LU and FT workloads from NAS parallel benchmarks \cite{bib-naspb2}. All of these benchmarks are parallelized using OpenMP and are compiled using \textit{-fopenmp} option of GCC. We use the OMP parameters to implement our methodology mentioned in the above section.

\begin{table}[h]
\centering
\begin{tabular}{|l|l|l|}
\hline
\textbf{Parameter} & \textbf{Description} & \textbf{Values} \\ \hline
OMP\_NUM\_THREADS & No. of threads & 1 to 16 \\ \hline
OMP\_PROC\_BIND & Enable/disable thread affinity & true/false \\ \hline
OMP\_PLACES & Specify availble cores & \{0\}..\{23\} \\ \hline
\end{tabular}
\caption{OpenMP Environment Variables}
\label{tab:setup1}
\end{table}

Table \ref{tab:setup1} describes the OpenMP environment variables and all the values for those variables. Here, we are using OMP\_PLACES environment variable to specify the exact core ids that the parallel application will use. This is important because this is how we achieve the effect of disabling hyper-threading. By setting the core list "\{0\},\{2\},\{4\},\{6\},\{8\},\{10\},\{12\},\{14\},
\{16\},\{17\},\{18\},\{19\},\{20\},\{21\},\{22\},\{23\}" to the OMP\_PLACES environment vairable before executing the application we skip all the logical (hyper-thread) cores and effectively disable hyper-threading for our workloads. The OMP\_NUM\_THREADS describes the no. of threads that are available to the application and the default no. of available threads is equal to the no. of available cores. Since in our case we 16 cores in our core list, we scale no. of threads from 1 to 16 in the increments of 2. The OMP\_PROC\_BIND variable is used to control the thread affinity of an application. When it is set to true, the thread scheduler will pin the threads to the available cores and avoids rescheduling them as much as possible. However, when it is set to false the thread scheduler is free to reschedule/move the threads as it sees fit. 

\subsection{Evaluation Results}

We see interesting scaling behaviour in the case of miniVite and FT. We can see from Fig \ref{fig:miniVite-true} where the experiment is run with OMP\_PROC\_BIND set to true, scaling across the P-/E-cores result in better scaling till 8 threads, but gets worse from 10 threads onto 16 threads. This can be explained by the fact that going from 8 to 10 threads implies that we are crossing of P-/E-core boundary and crossing that boundary results in threads being placed on E-cores which are only 60\% the speed of P-cores as mentioned in our earlier sections. However, a very peculiar and interesting observation here is that Fig \ref{fig:miniVite-false} shows that this effect of worse scaling is nullified when OMP\_PROC\_BIND set to false and the result is that we end up with better scaling. This is a major discovery in our research efforts and we carry on to try and explain why such an effect exists.

\begin{figure}[h]
     \centering
     \begin{subfigure}[b]{0.24\textwidth}
         \centering
         \includegraphics[width=\textwidth]{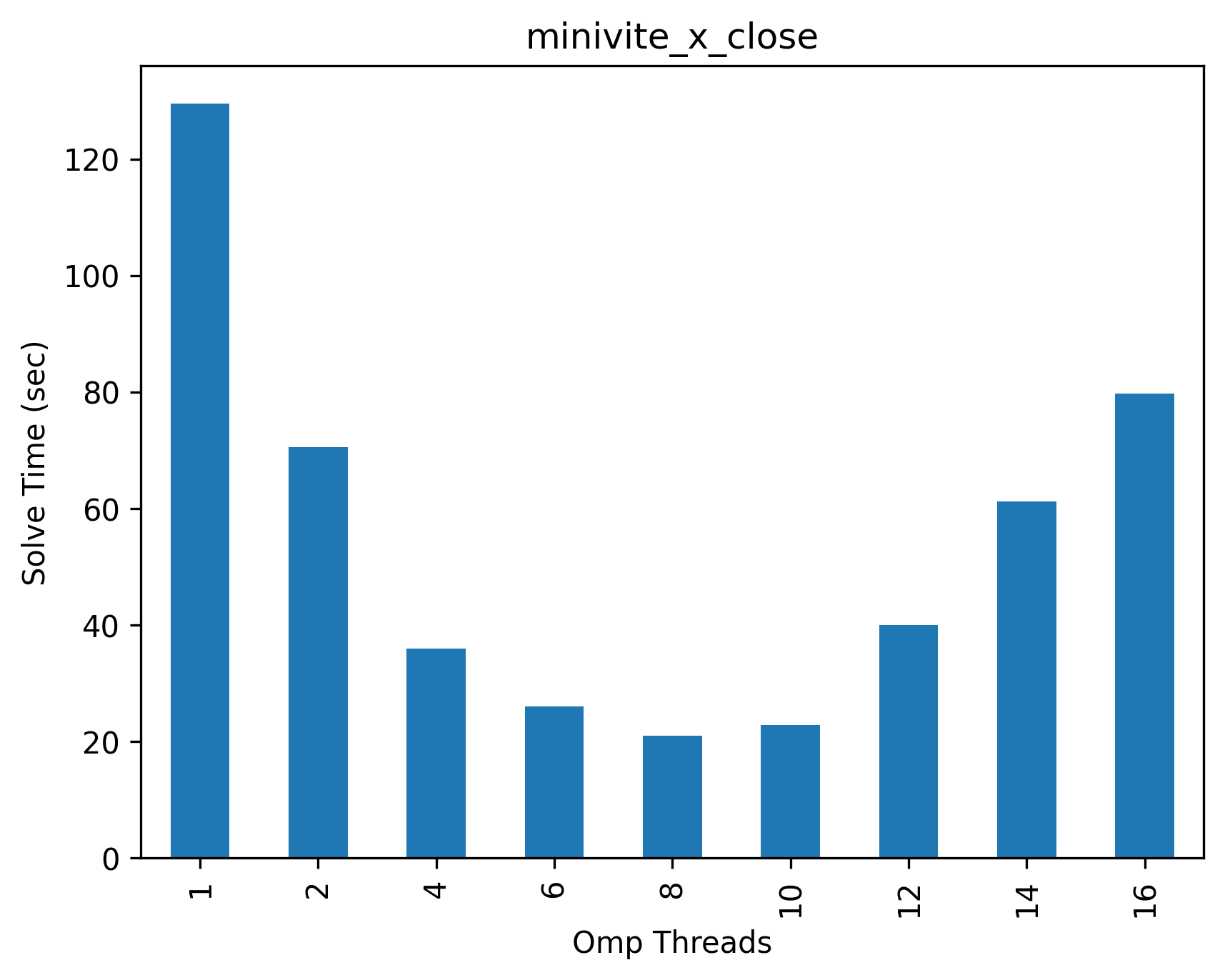}
         \caption{OMP\_PROC\_BIND=true}
         \label{fig:miniVite-true}
     \end{subfigure}
     \hfill
     \begin{subfigure}[b]{0.24\textwidth}
         \centering
         \includegraphics[width=\textwidth]{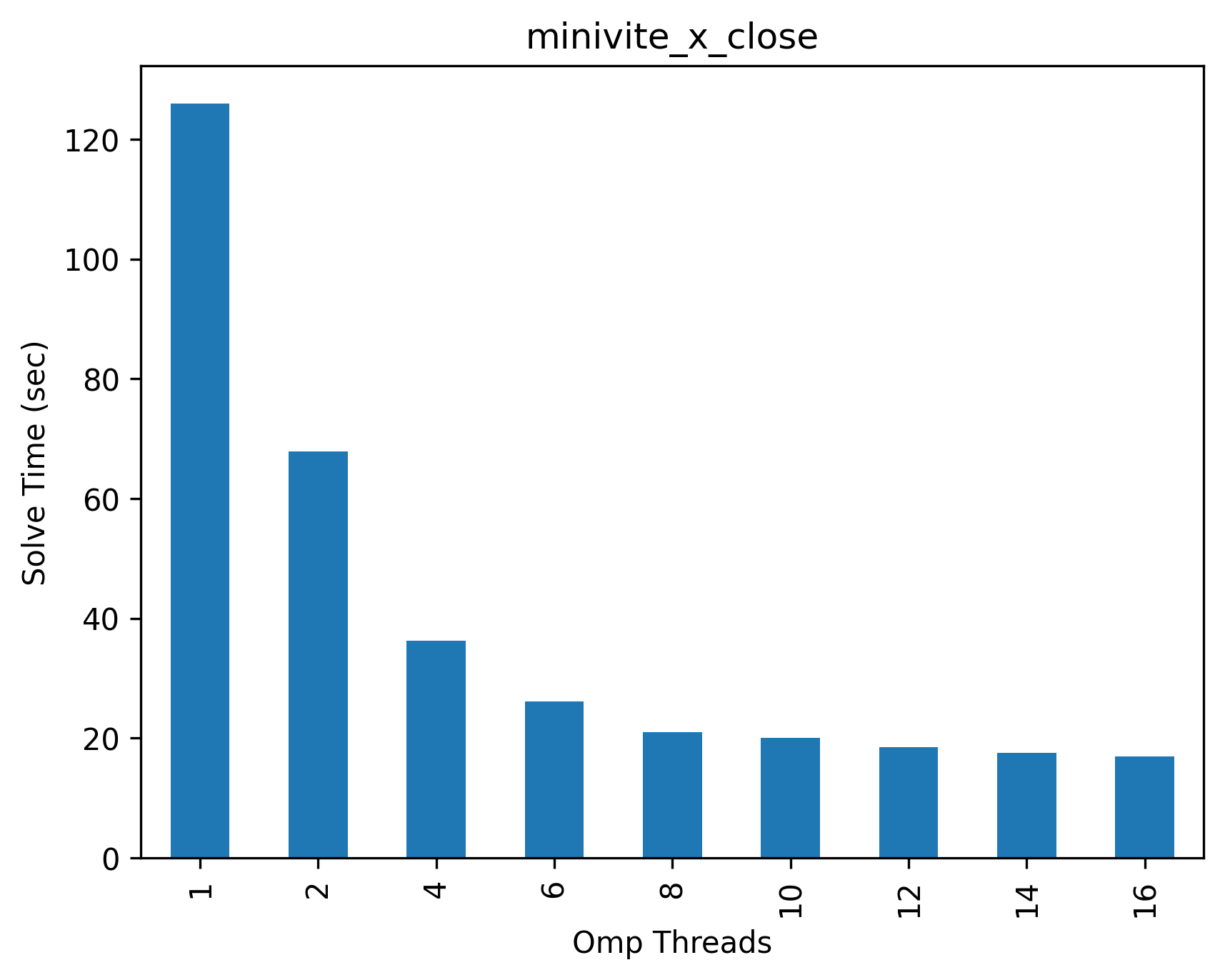}
         \caption{OMP\_PROC\_BIND=false}
         \label{fig:miniVite-false}
     \end{subfigure}
        \caption{\textit{miniVite}: Illustrating the impact of hybrid cores and thread affinity configurations}
        \label{fig:miniVite}
\end{figure}

Fig \ref{fig:ft} shows that this effect is not unique to just miniVite workload, but also can be noticed in the FT workload from NAS parallel benchmark. However, unlike miniVite, in this workload with thread affinty enabled, the scaling gets worse after crossing the P-core/E-core boundary only till 12 threads and trends down from 14 to 16 threads. Despite of that fact, the scaling gets better (similar to miniVite) when thread affinity is turned off and threads are allowed to move across cores.

\begin{figure}[h]
     \centering
     \begin{subfigure}[b]{0.24\textwidth}
         \centering
         \includegraphics[width=\textwidth]{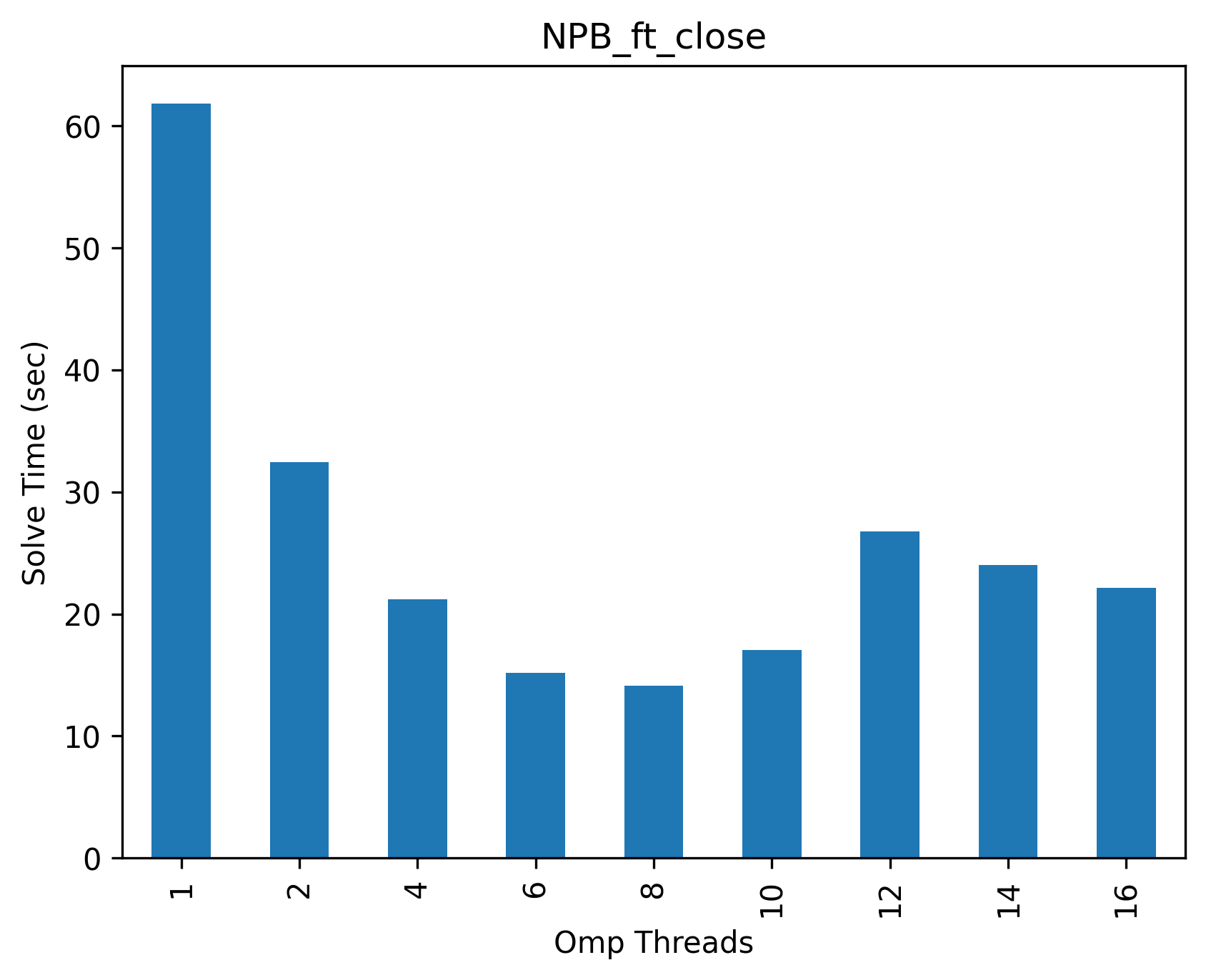}
         \caption{OMP\_PROC\_BIND=true}
         \label{fig:ft-true}
     \end{subfigure}
     \hfill
     \begin{subfigure}[b]{0.24\textwidth}
         \centering
         \includegraphics[width=\textwidth]{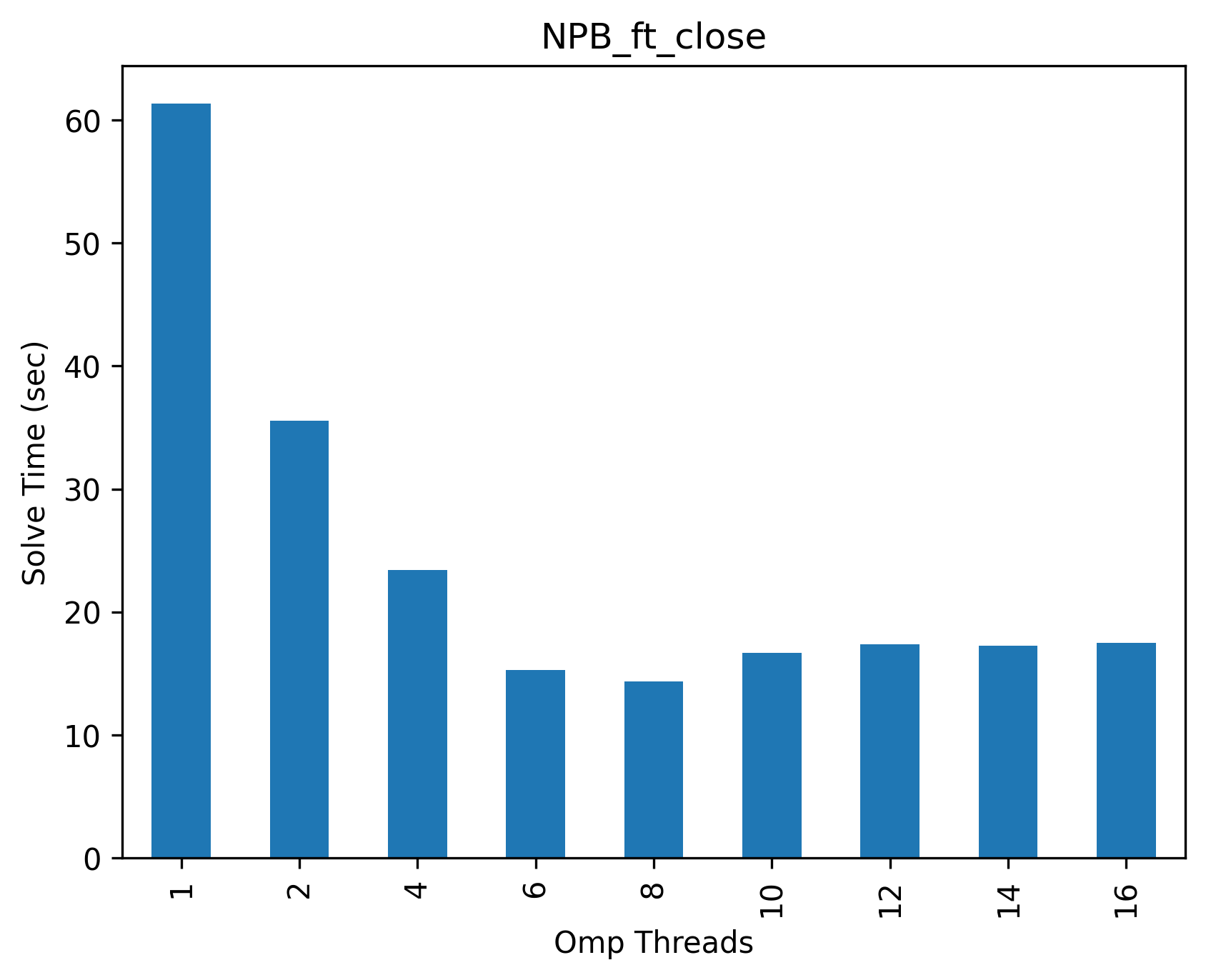}
         \caption{OMP\_PROC\_BIND=false}
         \label{fig:ft-false}
     \end{subfigure}
        \caption{NAS-FT: Illustrating }
        \label{fig:ft}
\end{figure}

\subsection{Unaffected workloads}
In our evaluation, not all workloads show this scaling behaviour. In fact, some workloads are unaffected by the existence of P-/E-core boundary. Moreover, turning the thread affinity on and off also has no effect on the scaling of such workloads.

\begin{figure}[h]
     \centering
     \begin{subfigure}[b]{0.24\textwidth}
         \centering
         \includegraphics[width=\textwidth]{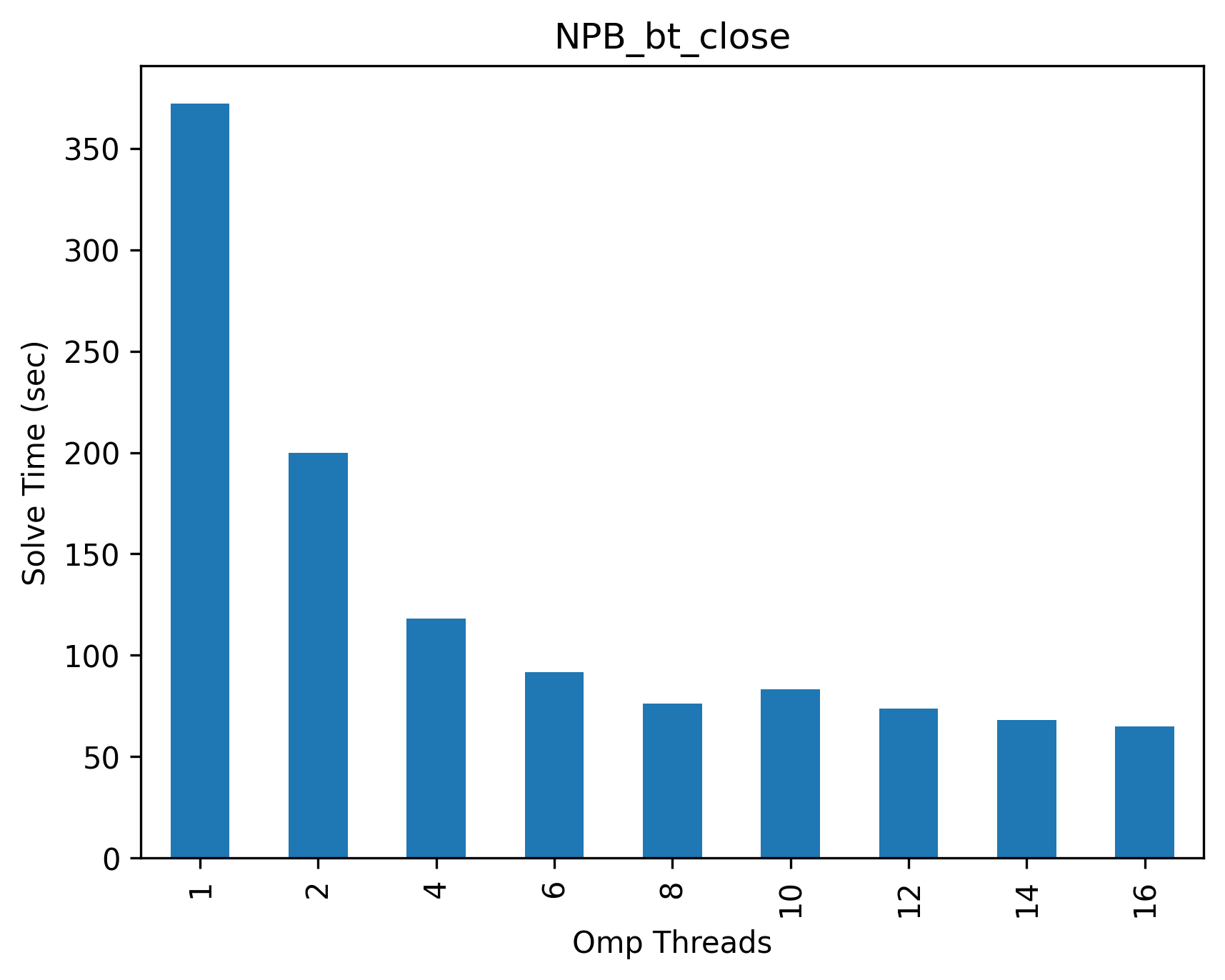}
         \caption{OMP\_PROC\_BIND=true}
         \label{fig:bt-true}
     \end{subfigure}
     \hfill
     \begin{subfigure}[b]{0.24\textwidth}
         \centering
         \includegraphics[width=\textwidth]{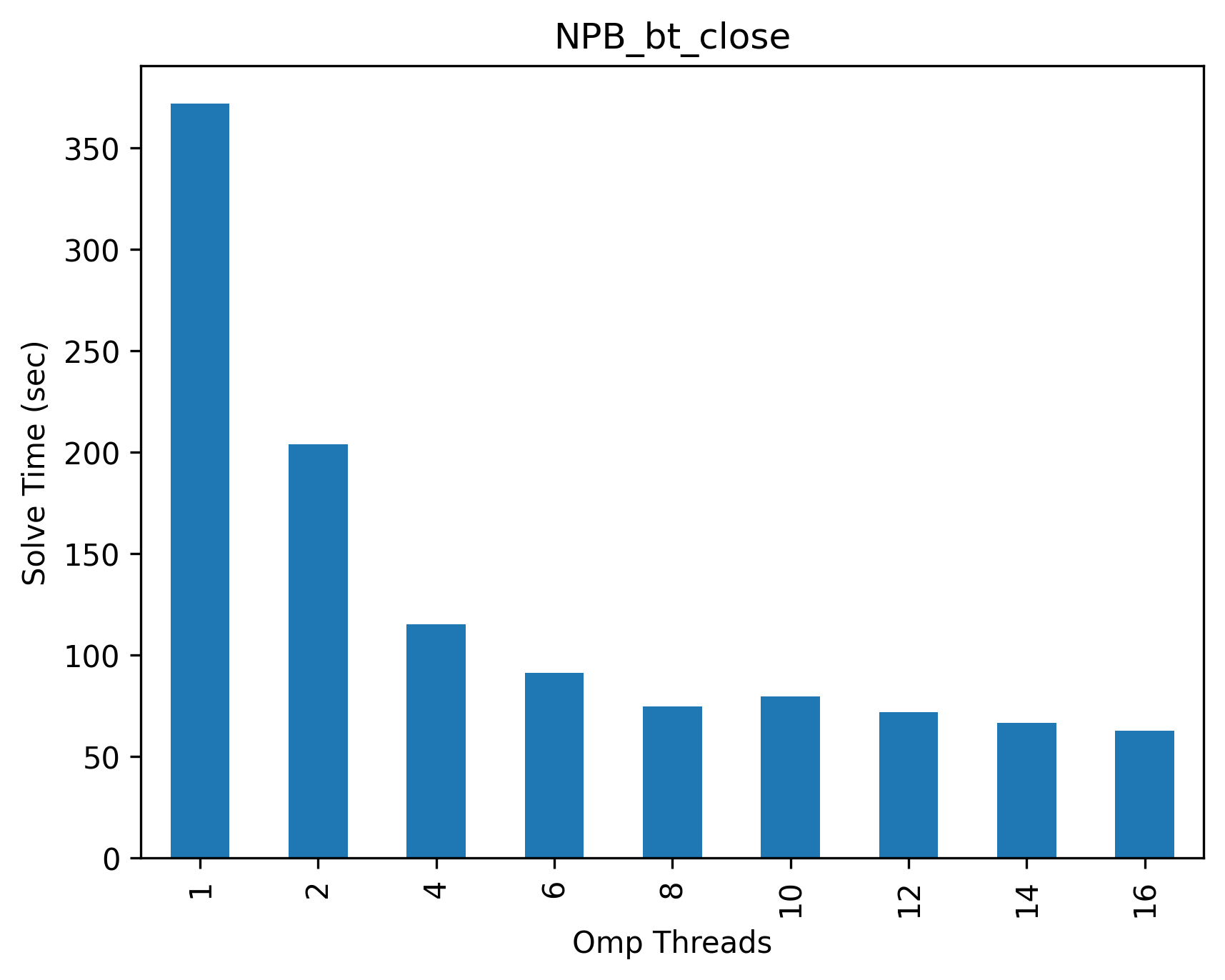}
         \caption{OMP\_PROC\_BIND=false}
         \label{fig:bt-false}
     \end{subfigure}
        \caption{NAS-BT: Unaffected by the hybrid cores and thread affinity configurations}
        \label{fig:bt}
\end{figure}

\begin{figure}[h]
     \centering
     \begin{subfigure}[b]{0.24\textwidth}
         \centering
         \includegraphics[width=\textwidth]{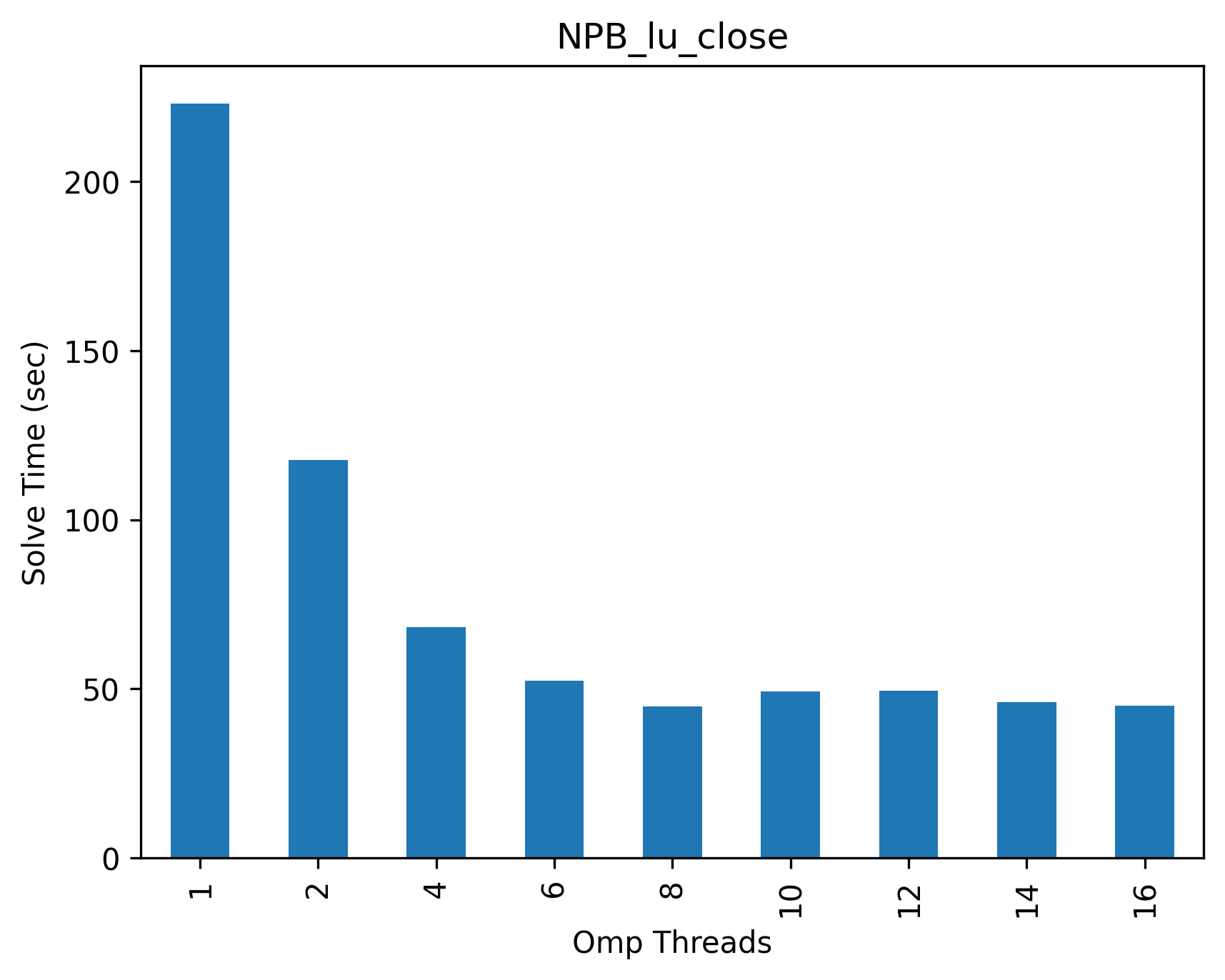}
         \caption{OMP\_PROC\_BIND=true}
         \label{fig:y lu-true}
     \end{subfigure}
     \hfill
     \begin{subfigure}[b]{0.24\textwidth}
         \centering
         \includegraphics[width=\textwidth]{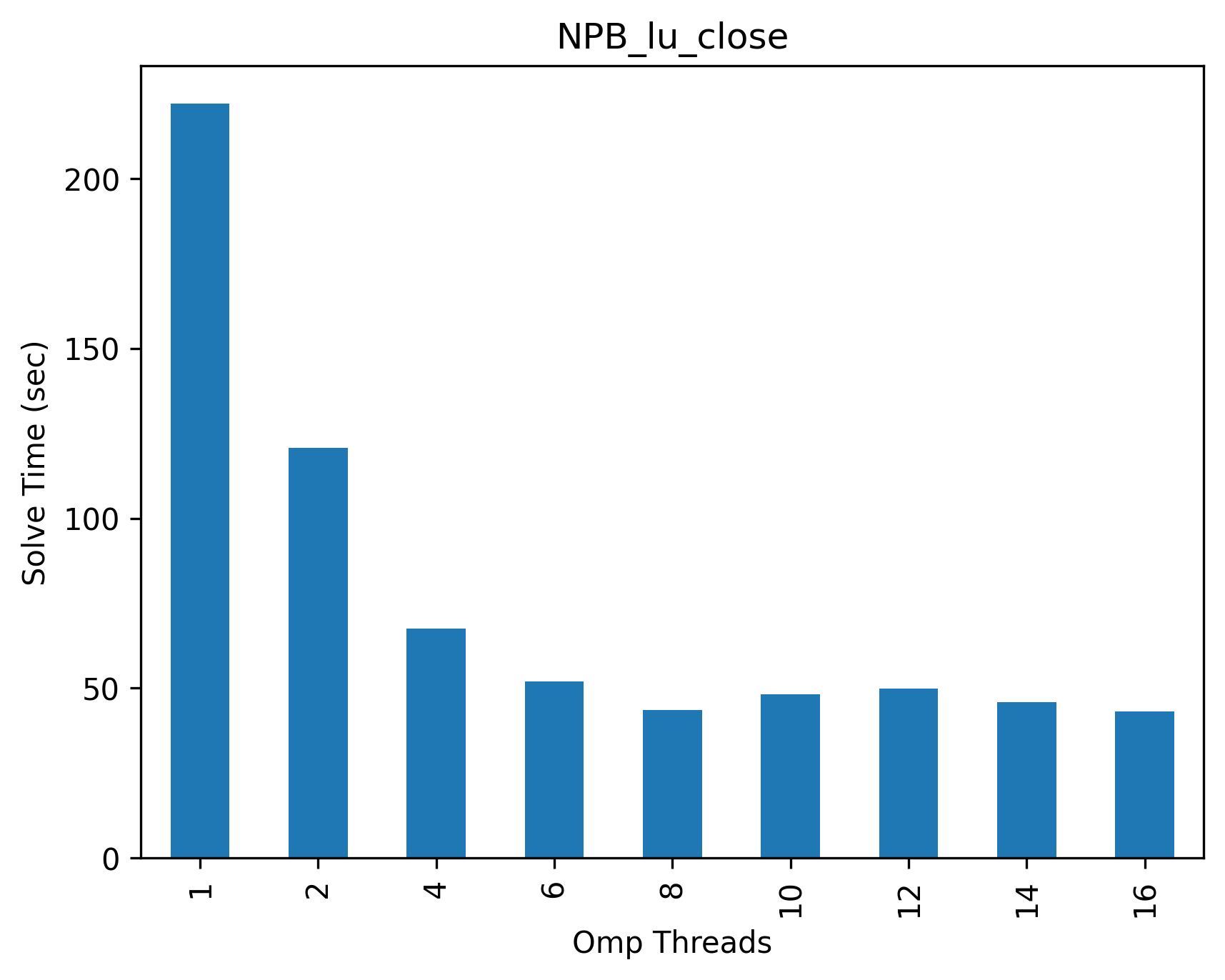}
         \caption{OMP\_PROC\_BIND=false}
         \label{fig:lu-false}
     \end{subfigure}
        \caption{NAS-LU: Unaffected by the hybrid cores and thread affinity configurations}
        \label{fig:lu}
\end{figure}

\begin{figure}[h]
     \centering
     \begin{subfigure}[b]{0.24\textwidth}
         \centering
         \includegraphics[width=\textwidth]{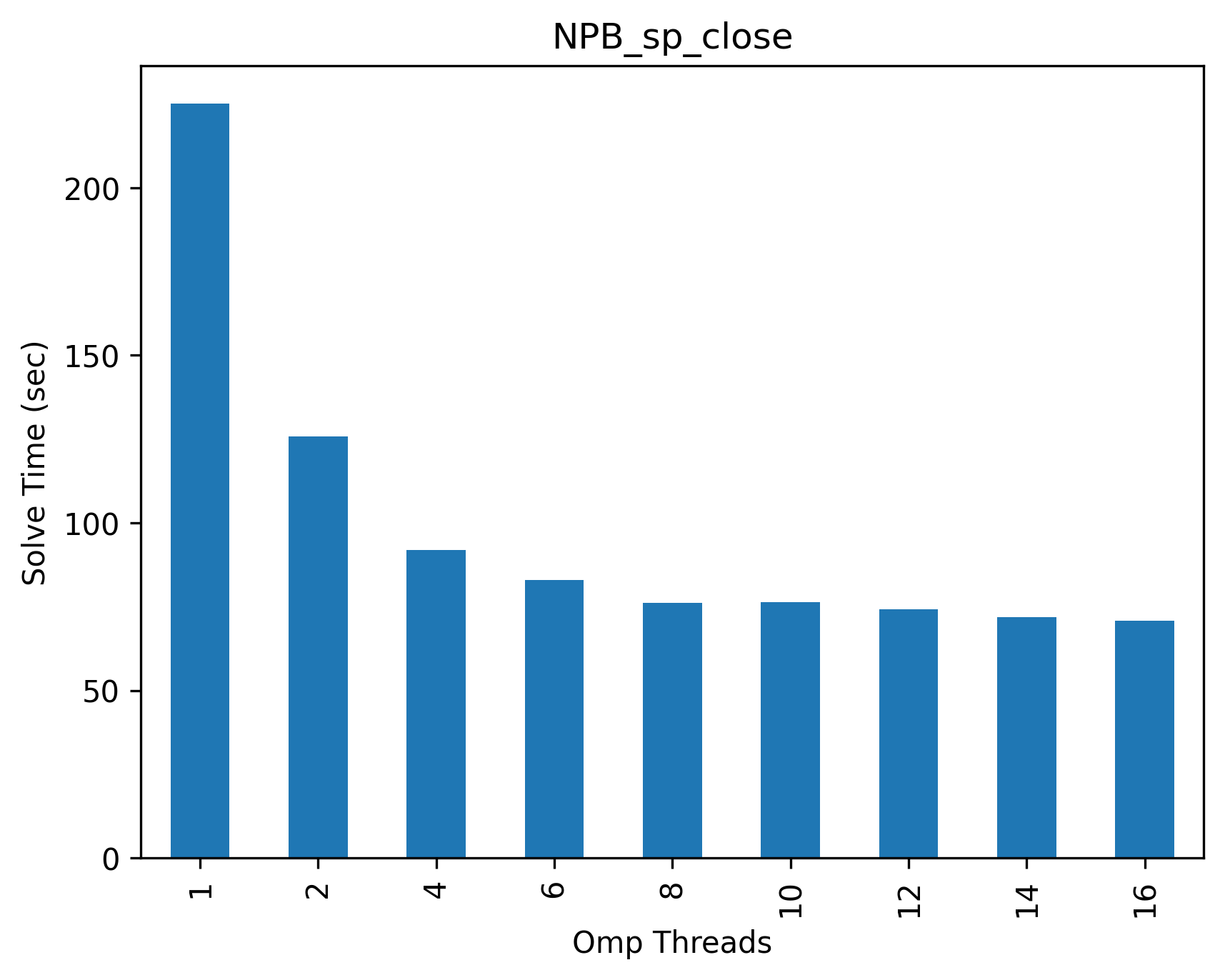}
         \caption{OMP\_PROC\_BIND=true}
         \label{fig:sp-true}
     \end{subfigure}
     \hfill
     \begin{subfigure}[b]{0.24\textwidth}
         \centering
         \includegraphics[width=\textwidth]{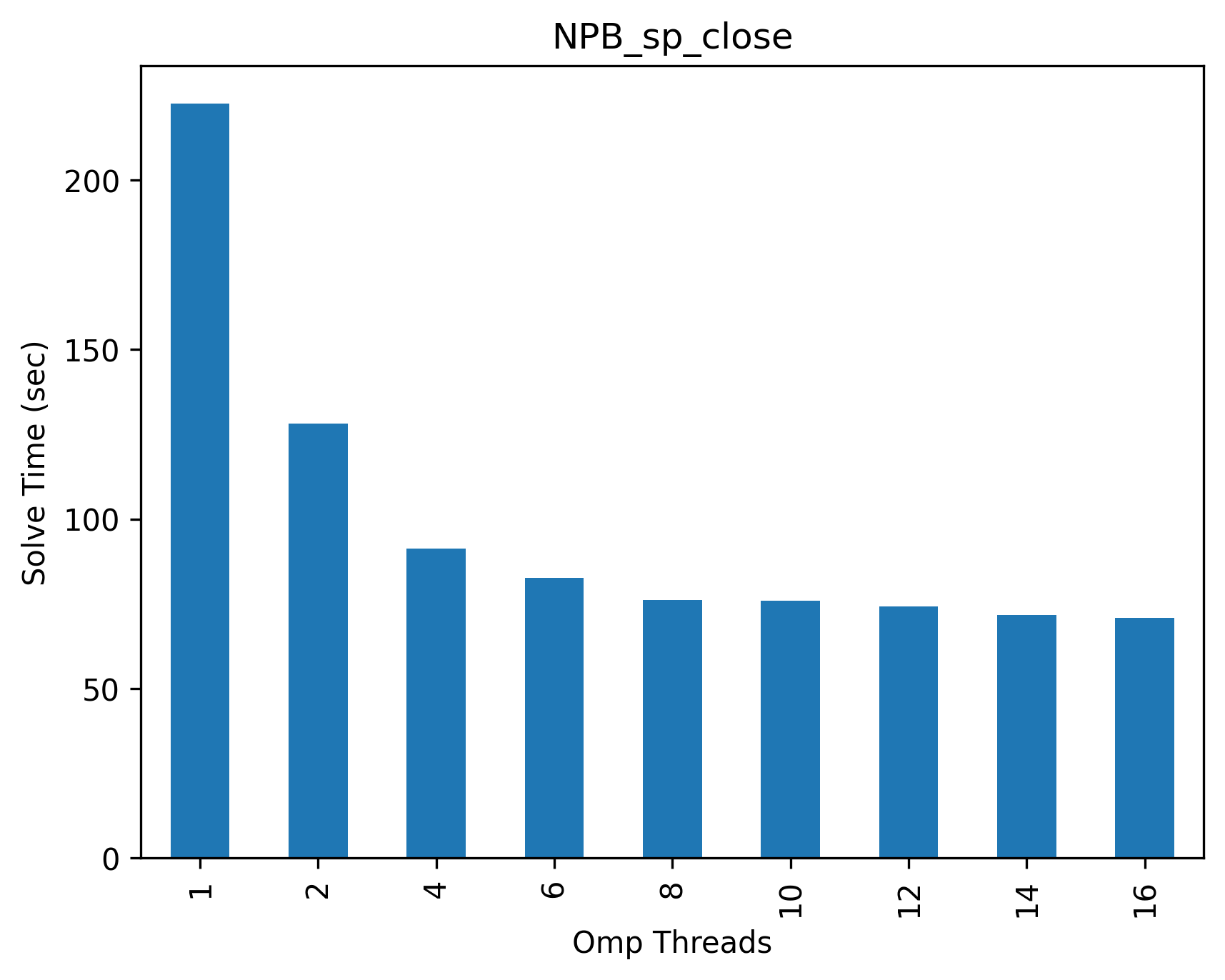}
         \caption{OMP\_PROC\_BIND=false}
         \label{fig:sp-false}
     \end{subfigure}
        \caption{NAS-SP: Unaffected by the hybrid cores and thread affinity configurations}
        \label{fig:sp}
\end{figure}

Figures \ref{fig:bt}, \ref{fig:lu}, and \ref{fig:sp} show BT, LU, and SP respectively from NAS parallel benchmarks that are unaffected by both P-core/E-core boundary and thread affinity configurations.

\section{Explanation of the Impact of Hybrid Cores}

To explain the phenomenon of such interesting scaling patterns for some parallel workloads, in our case miniVite and FT, we started with the hypothesis that this effect must be due to the memory access patterns of these applications which in turn results in work imbalance among parallel threads. This is a reasonable hypothesis because miniVite is a graph workload that involves a lot complex pointer chasing in memory (it changes based on input graph) and FT is a Fourier Transform workloads with butterfly memory access pattern. We investigate this by collecting memory traces and looking at the per-core memory footprints of the parallel workload under consideration (miniVite) and qualitatively come to the conclusion that work imbalance in threads lead to better scaling across hybrid cores given that the thread affinity is disabled.

\subsection{Collecting Memory Footprint Using MemGaze}

The idea here is to look at the memory access of the miniVite or FT to determine whether there is a relationship between memory access pattern of a workload and the thread scaling behaviour. To measure the memory access patterns we used a recently publish tool called MemGaze \cite{bib-memgaze}. MemGaze is a lightweight memory analysis tool that uses \textit{perf} with PT\_WRITE support to rapidly collect memory traces of an application and it performs in-depth memory analysis on those traces giving result to various metrics about the memory accesses of an application such as: foot-print, data-reuse etc. Here we focus on the memory footprint metric to understand the relationship between memory accesses and scaling behaviour. Memory \textit{footprint} is defined as the no. of unique memory accesses that an application invokes during it's execution.
  
However, when we started this effort MemGaze didn't support analyzing memory accesses per-core at the time. So, I have modified MemGaze to add support for per-core memory footprint calculations. After which, we used MemGaze tool to collect the per-core memory footprint and analyze the per-core memory access patterns.

We focus on collecting per-core memory traces for \textit{miniVite}. However, collecting all the traces should be avoided because the resulting trace file can easily exceed the system primary and secondary memory resources. Thus, we use MemGaze's sampling input and focus function input to collect the per-core memory footprints of major parallel section of \textit{miniVite} workload. 

\begin{figure*}[!t]
     \centering
     \begin{subfigure}[b]{0.45\textwidth}
         \centering
         \includegraphics[width=\textwidth]{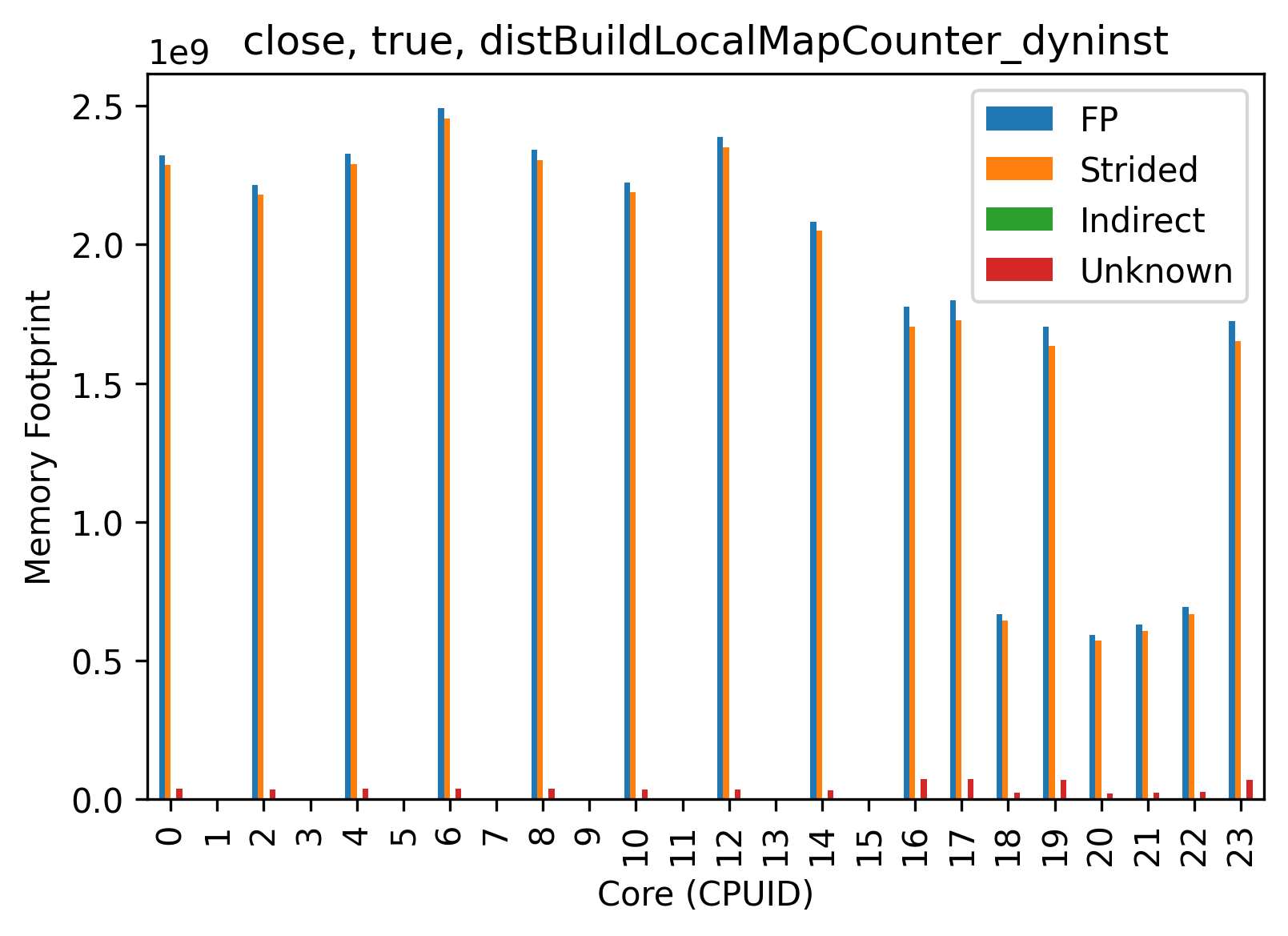}
         \caption{OMP\_PROC\_BIND=true}
         \label{fig:cpufp-true}
     \end{subfigure}
     \hfill
     \begin{subfigure}[b]{0.45\textwidth}
         \centering
         \includegraphics[width=\textwidth]{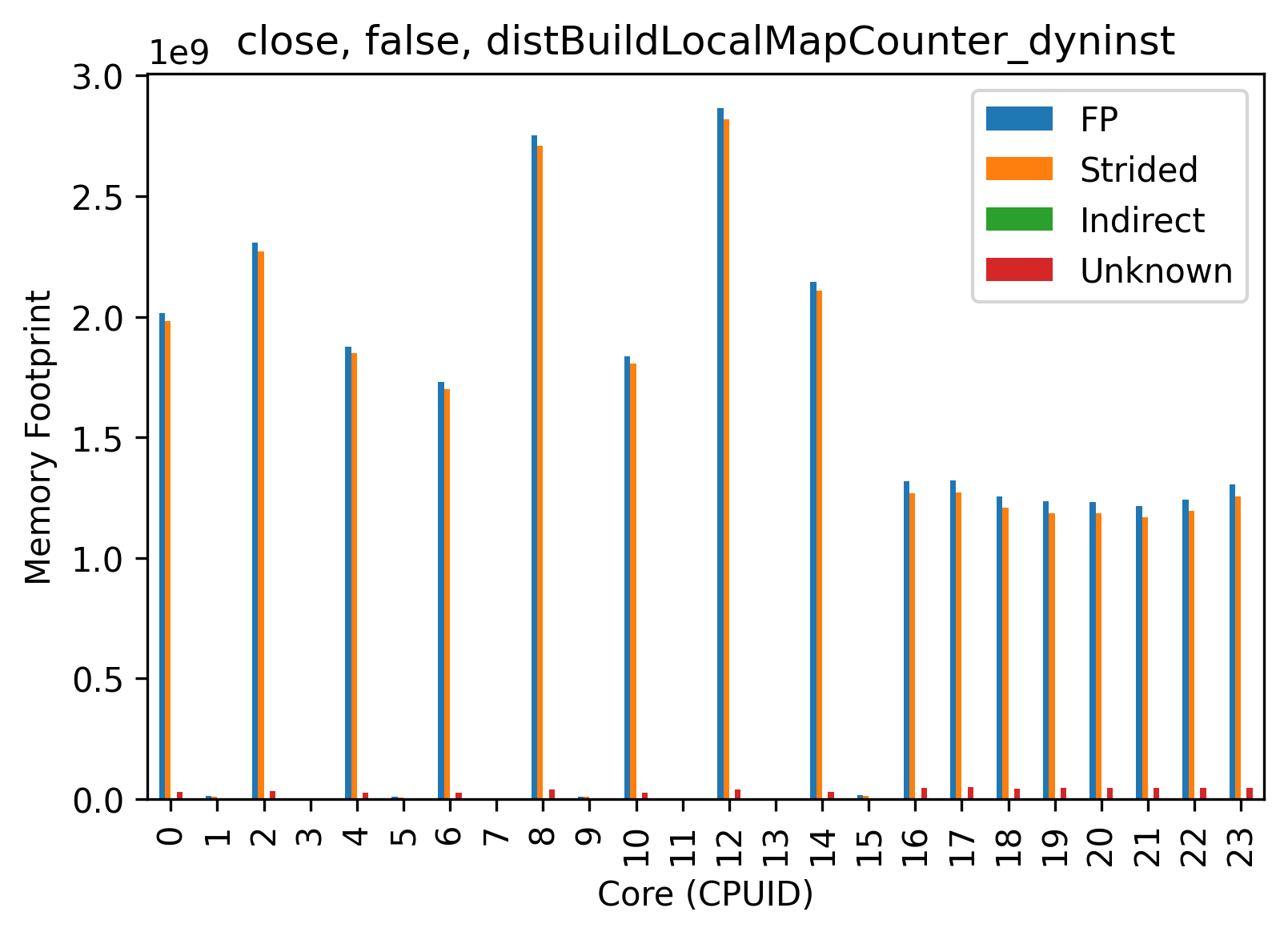}
         \caption{OMP\_PROC\_BIND=false}
         \label{fig:cpufp-false}
     \end{subfigure}
        \caption{Per-core memory footprint of the parallelized OMP function where no. of threads is set to 16, focus function is set to \textit{distBuildLocalMapCounter\_dyinst} and \ref{fig:cpufp-true}, \ref{fig:cpufp-false} are the cases where thread affinity is set to true, false respectively. The resulting output has FP (footprint: total unique accesses), Strited (unique strided accesses), Indirect (unique indirect accesses), Unknown (unrecognized/unresolved memory accesses) }
        \label{fig:cpufp}
\end{figure*}

\subsection{Work Imbalance Leads to Better Scaling} 

Fig \ref{fig:cpufp} shows the difference between the per-core memory footprints of OMP parallelized function \textit{distBuildLocalMapCounter\_dyinst} when thread affinity is set to true vs. false. 

As we can see, the max per-core memory footprint is larger in the case where thread affinity is disabled, which implies that there is more activity on some cores when in this case compared to the case where thread affinity is enabled. Another important observation here is that, on E-cores, the footprints are not uniform in the case of \ref{fig:cpufp-true} and are uniform in case of \ref{fig:cpufp-false}. This implies that, turning of the affinity results in more uniform activity on E-cores. This can be true because if the affinity is set to false then the thread scheduler is free to move threads across cores which can resulting in such access patterns. To verify this, we measure the no. of thread migrations that occur during the execution of \textit{miniVite} using \textit{perf}. Fig \ref{fig:migrations} shows that there are more thread migrations when affinity is turned off (false) when compared to the case when it is turned on (true). 

\begin{figure}[h]
    \centering
     \includegraphics[width=0.45\textwidth]{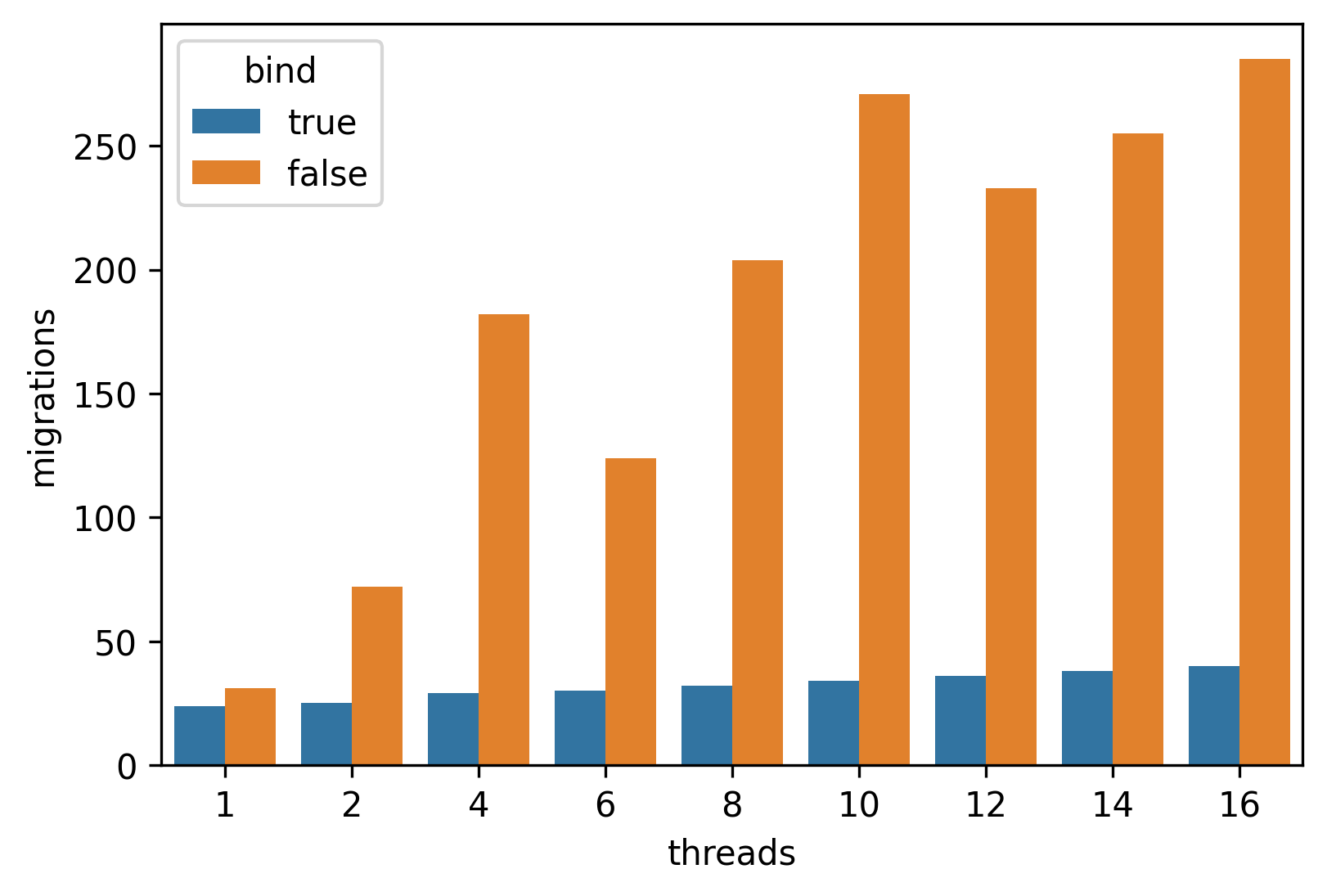}
     \caption{Thread migrations of \textit{miniVite} that are observed using \textit{perf} by settings OMP\_PROC\_BIND set to true vs. false}
     \label{fig:migrations}
\end{figure}

The difference in thread activity on cores illustrated in Fig \ref{fig:cpufp} can be characterized as the existence of \textit{work imbalance}, i.e., not all threads in a workload are doing same amount of work. This work imbalance in parallel applications implies inefficient usage of parallel compute resources (cores) when the threads are pinned to cores, i.e., in the case where a thread with more work to do than another one and both are pinned to core, then one of the core is going be idle (assuming it's only application that is running the system, which is often the case while running HPC workloads). However if the thread affinity is turned off then scheduler naturally sees the imbalances in work and imbalance in core frequencies (P-cores vs. E-cores) and tries to match the work with the available resources more efficiently given the cost of migrating a thread less than the time wasted during idle computation. Thus explaining the uniform distribution of footprint in Fig \ref{fig:cpufp-false}. By contrast, the same effect won't hold if all the cores are homogeneous (same frequency), because the scheduler doesn't have an incentive to migrate a thread to different core as all cores have similar capability and the cost of migration would just be an overhead.

Thus with this reasoning we can qualitatively say that the applications with work imbalance can benefit from the existence of hybrid cores and by turning off the thread affinity to allow the thread scheduler to find efficient thread placements dynamically.

\section{Aspect 2: Impact of Hybrid L2 Cache}

As mentioned in Section II, this new Alderlake CPU, along with hybrid core architecture, has a hybrid cache architecture with partially shared L2 cache among E-cores. Thus prompting to asking the question: how does this hybrid cache architecture effect the performance of parallel applications?

\subsection{Methodology}

The methodology for this approach is very similar to the one followed in Section IV, except the only difference here is that we are not changing thread affinity (set to true) and controlling the available cores by setting OMP\_PLACES. The basic idea of this aspect is to run parallel workloads that have spatially and temporally local concurrent memory accesses by \textit{co-locating (l2-shared)} threads on E-cores that share same L2 and comparing it against the scenario where the same threads are \textit{not colocated (l2-split)}. 

The hypothesis here is that, since L2 cache is partially split across E-cores, the performance of parallel workloads (with spatially and temporally local concurrent memory accesses) should perform better in case of \textit{l2-shared}
 when compared to the case of \textit{l2-split}
 
\subsection{Test Setup}

To test the L2 cache impact on performance, we chose 2 thread configurations (2, 4) and 2 thread placement configurations (\textit{l2-shared, l2-split}), which are permuted (2x2=4) and detailed in Table \ref{tab:setup2}.

\begin{table}[h]
\centering
\begin{tabular}{|l|l|l|}
\hline
\textbf{Configuration} & \textbf{Threads} & \textbf{OMP\_PLACES} \\ \hline
l2shared & 2 & {16},{17} \\ \hline
l2shared & 4 & {16},{17},{18},{19} \\ \hline
l2split & 2 & {16},{20} \\ \hline
l2split & 4 & {16},{17},{20},{21} \\ \hline
\end{tabular}
\caption{\textit{l2-shared} and \textit{l2-split} Configurations}
\label{tab:setup2}
\end{table}

We evaluate this setup by running number of benchmarks, starting with, \textit{BT, FT, LU and SP} from NAS Parallel Benchmarks to show that data parallel workloads show the opposite effect of our previous hypothesis. We then go on to evaluate selected workloads: \textit{blackscholes, bodytrack, dedup, streamcluster} from PARSEC \cite{bib-parsec1}, \cite{bib-parsec2} benchmark which are known to have shared cache effects \cite{bib-thread-cache}. Finally we evaluate workloads with atomic accesses and locks such as HiParTi \cite{bib-hiparti-git}, \cite{bib-hicoo} and CRONO \cite{bib-crono} workloads.

\subsection{Evaluation Results}

\begin{figure}[h]
     \centering
     \begin{subfigure}[b]{0.24\textwidth}
         \centering
         \includegraphics[width=\textwidth]{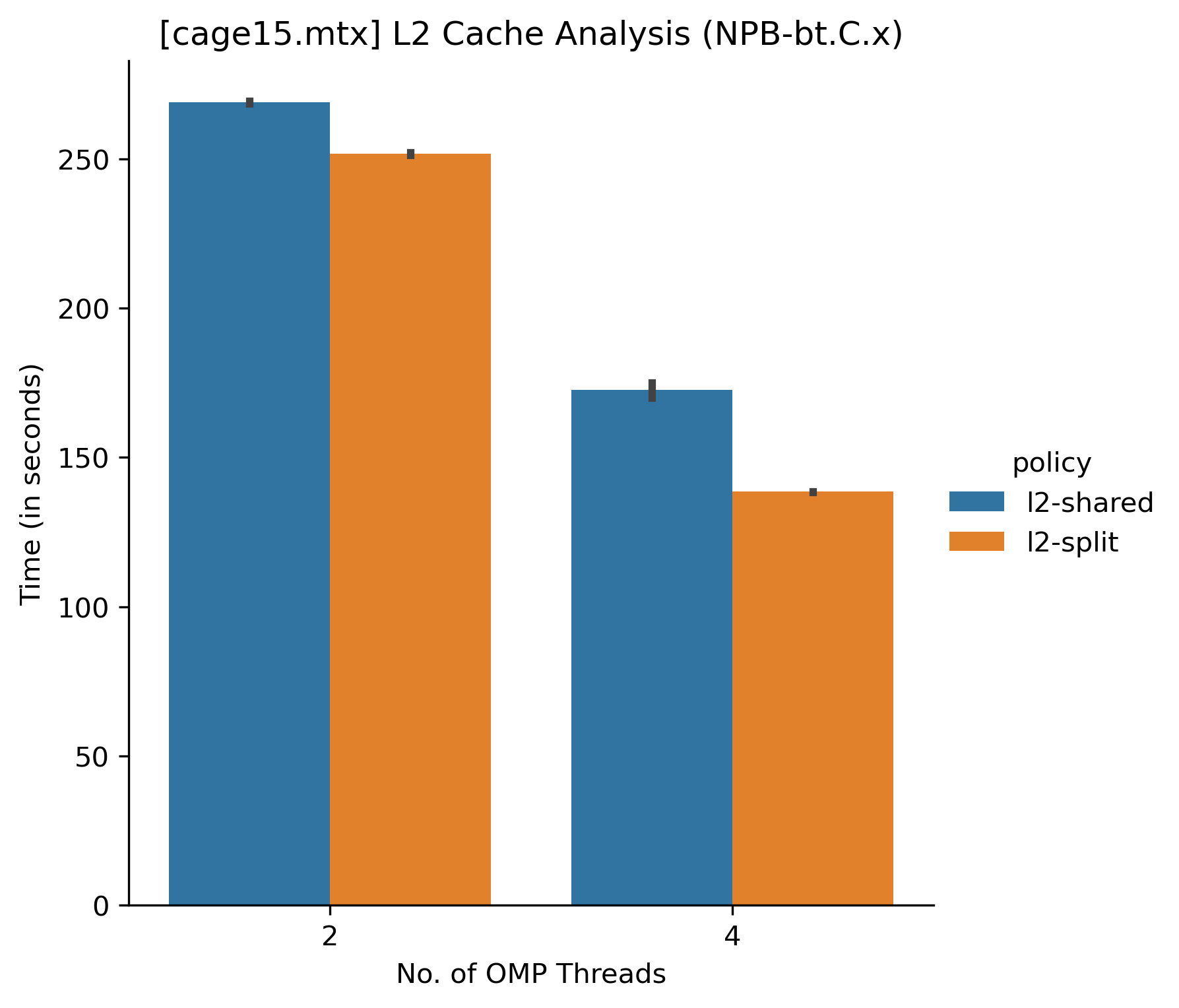}
         \caption{BT}
         \label{fig:l2-bt}
     \end{subfigure}
     \hfill
     \begin{subfigure}[b]{0.24\textwidth}
         \centering
         \includegraphics[width=\textwidth]{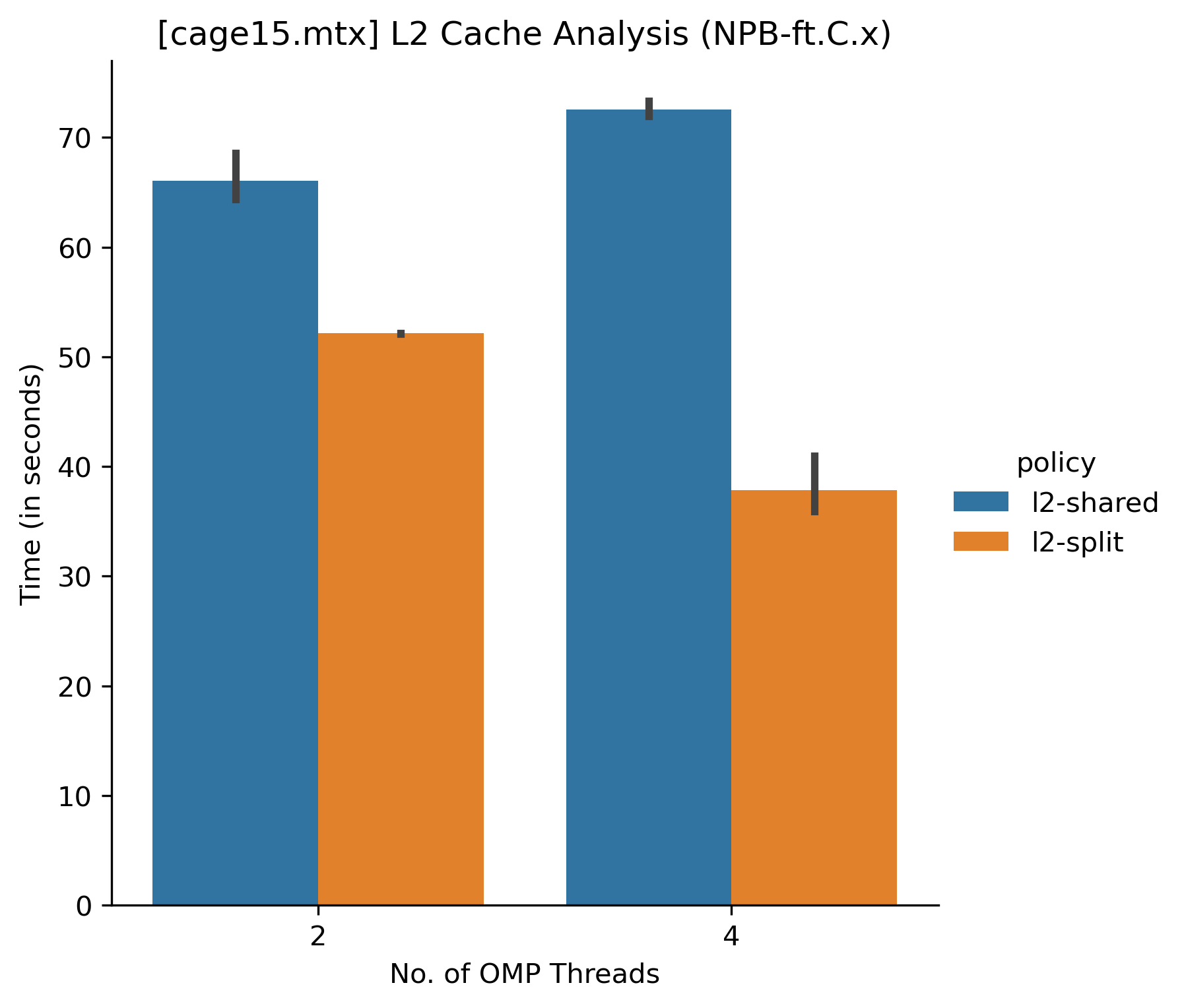}
         \caption{FT}
         \label{fig:l2-ft}
     \end{subfigure}
      \hfill
     \begin{subfigure}[b]{0.24\textwidth}
         \centering
         \includegraphics[width=\textwidth]{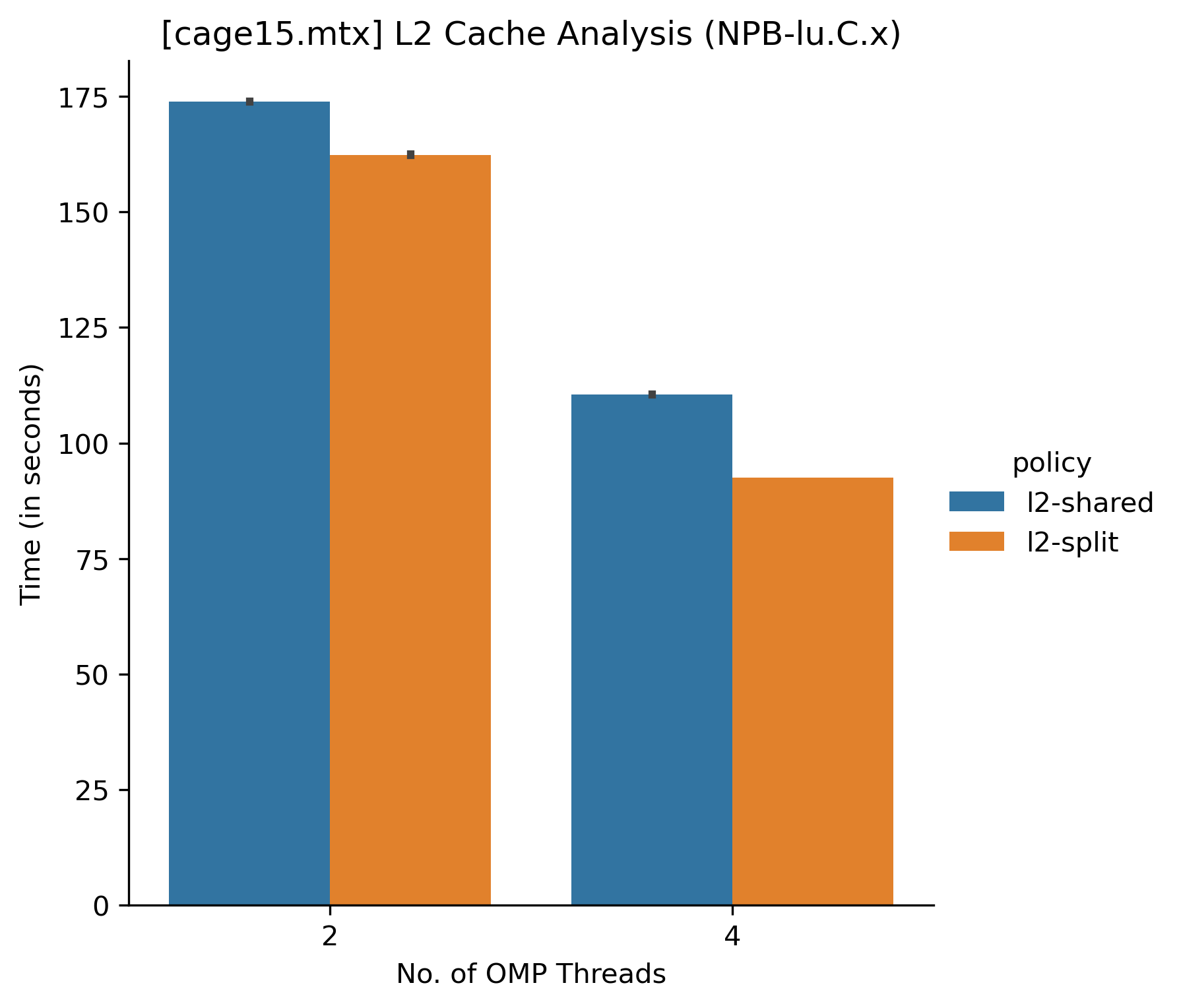}
         \caption{LU}
         \label{fig:l2-lu}
     \end{subfigure}
      \hfill
     \begin{subfigure}[b]{0.24\textwidth}
         \centering
         \includegraphics[width=\textwidth]{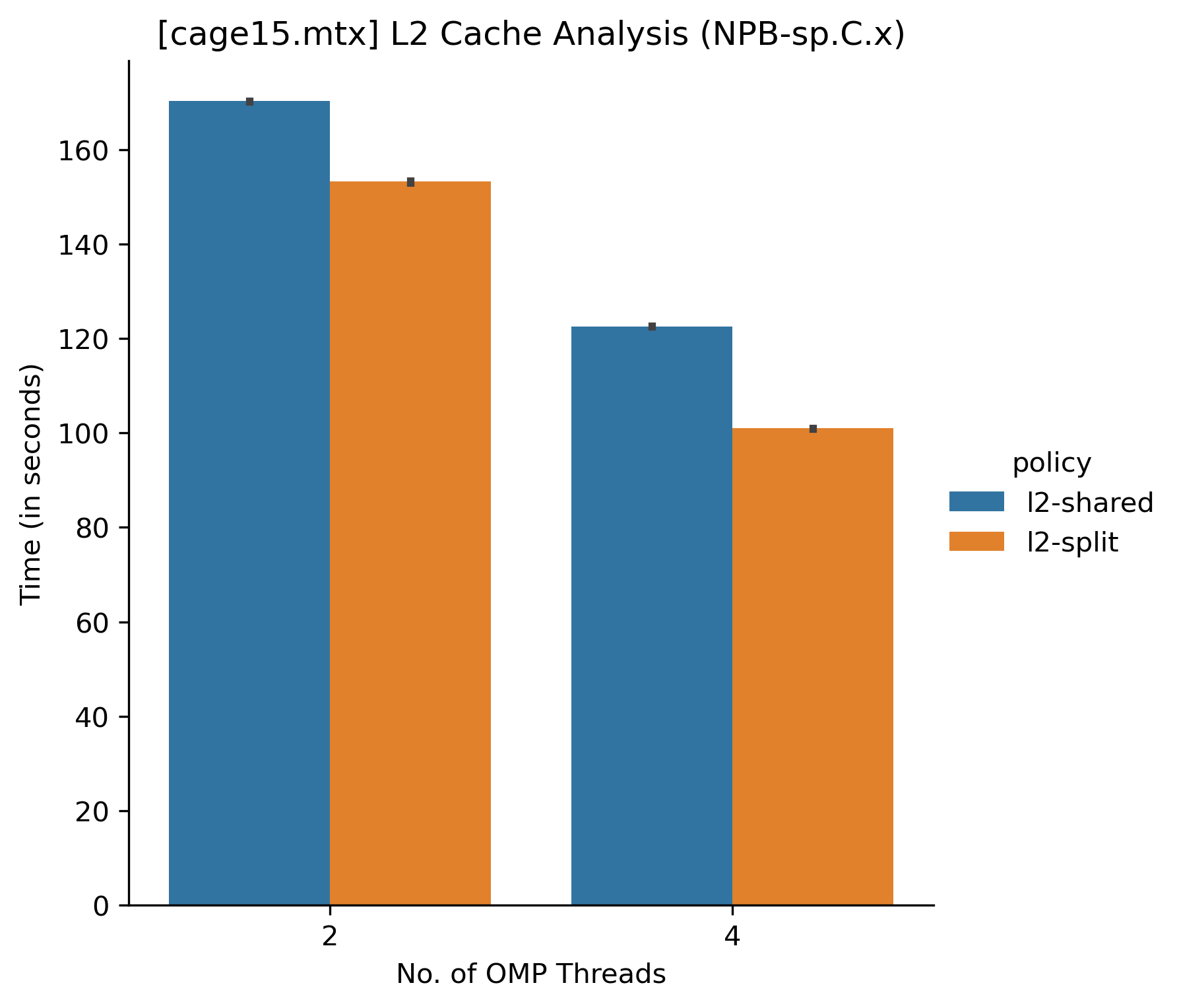}
         \caption{SP}
         \label{fig:l2-sp}
     \end{subfigure}
        \caption{L2 impact on NAS Parallel Benchmarks showing \textit{l2-split} placement is better than \textit{l2-shared} which is trivial given all these workloads are data-parallel}
        \label{fig:l2-nas}
\end{figure}

NAS parallel benchmarks show something that is trivial, that data-parallel workloads benefit from threads having more cache capacity. That is why, Fig \ref{fig:l2-nas} shows \textit{l2-split} thread placement with more cache capacity per thread perform better than \textit{l2-shared} thread placement even with scaling up the no. of threads.

\begin{figure}[h]
    \centering
     \includegraphics[width=0.45\textwidth]{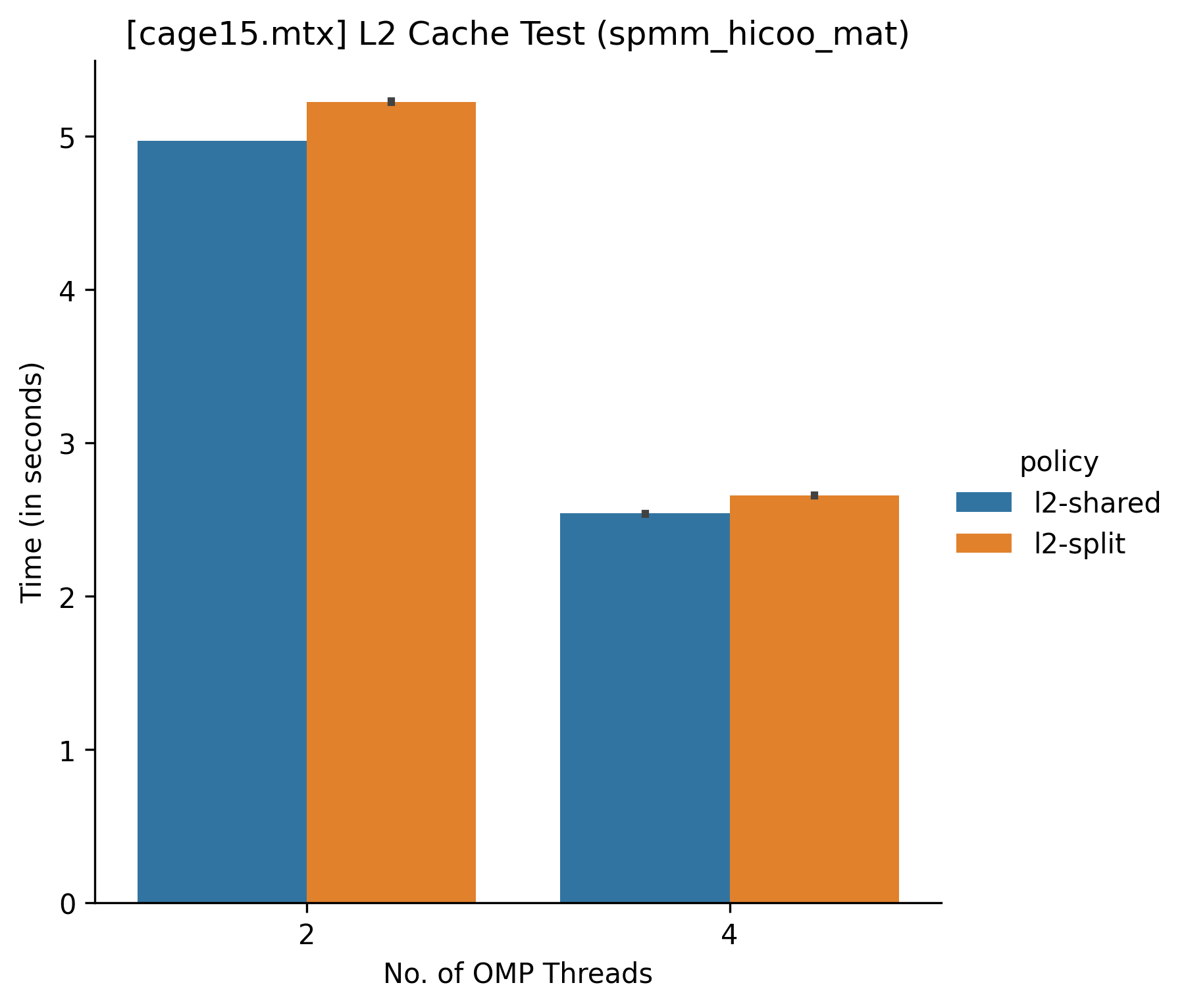}
     \caption{HiParTi: Sarse matrix multiplication workload with HiCOO matrix representation illustrating the impact of \textit{l2-shared} vs. \textit{l2-split}}
     \label{fig:l2-spmm-hicoo}
\end{figure}

HiParTi workload with atomic update on a shared data, shows minimal \textit{l2-shared} performance improvement compared to \textit{l2-split} which is illustrated in Fig \ref{fig:l2-spmm-hicoo}. This is consistent with the scaling up of no. of threads and data/workload size.

\begin{figure}[h]
     \centering
     \begin{subfigure}[b]{0.24\textwidth}
         \centering
         \includegraphics[width=\textwidth]{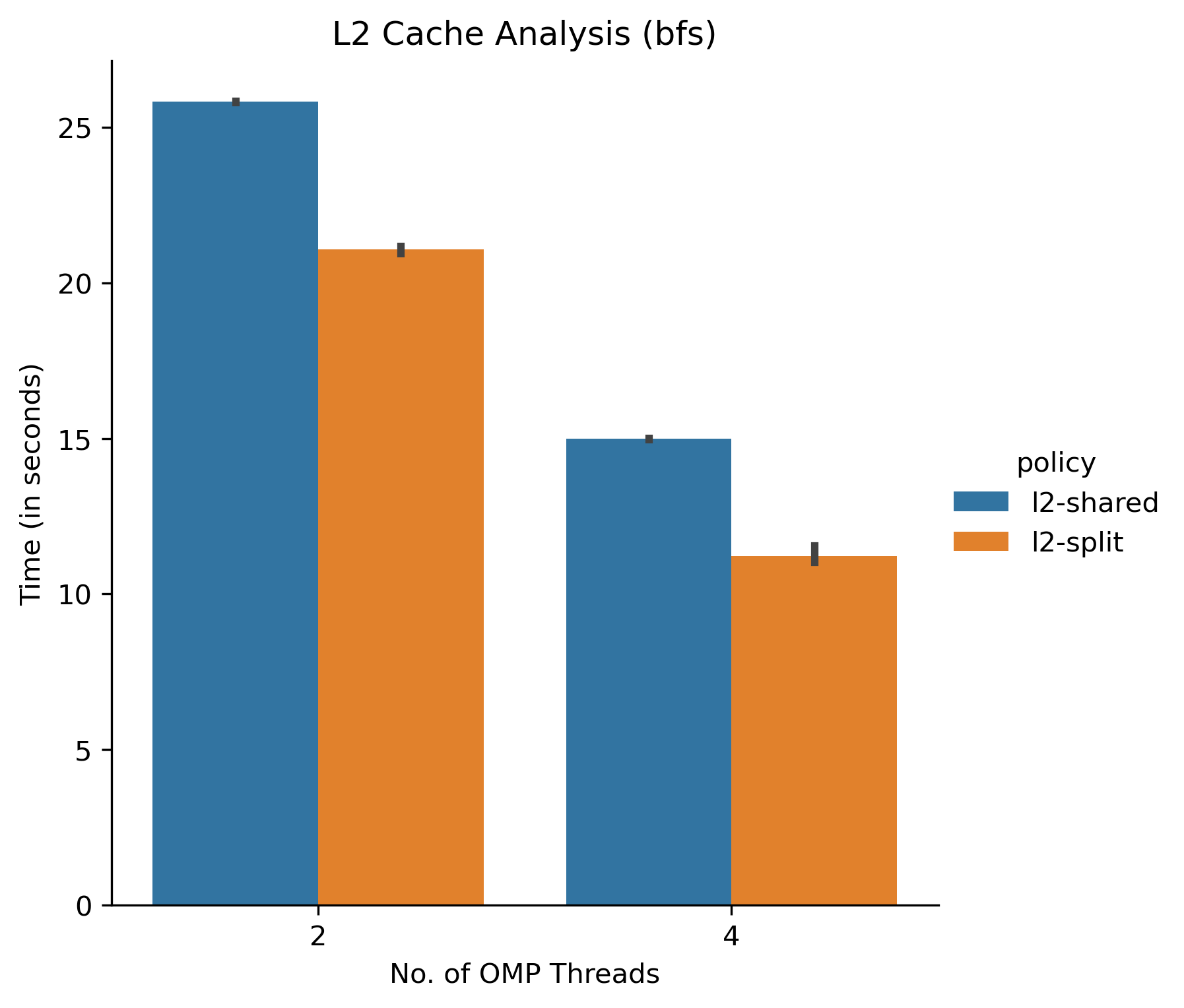}
         \caption{BFS}
         \label{fig:crono-bfs}
     \end{subfigure}
     \hfill
     \begin{subfigure}[b]{0.24\textwidth}
         \centering
         \includegraphics[width=\textwidth]{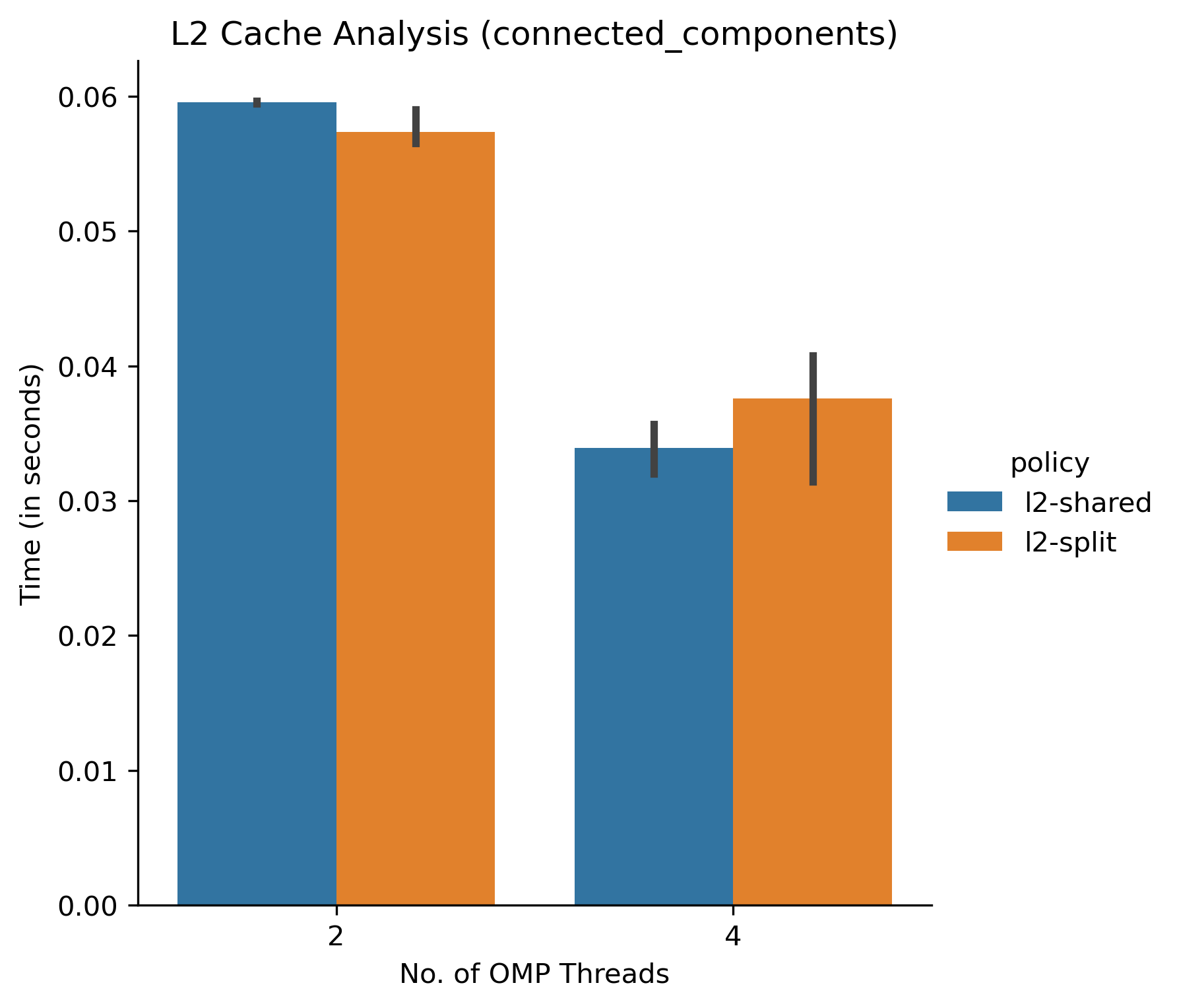}
         \caption{Connected Components}
         \label{fig:crono-cc}
     \end{subfigure}
      \hfill
     \begin{subfigure}[b]{0.24\textwidth}
         \centering
         \includegraphics[width=\textwidth]{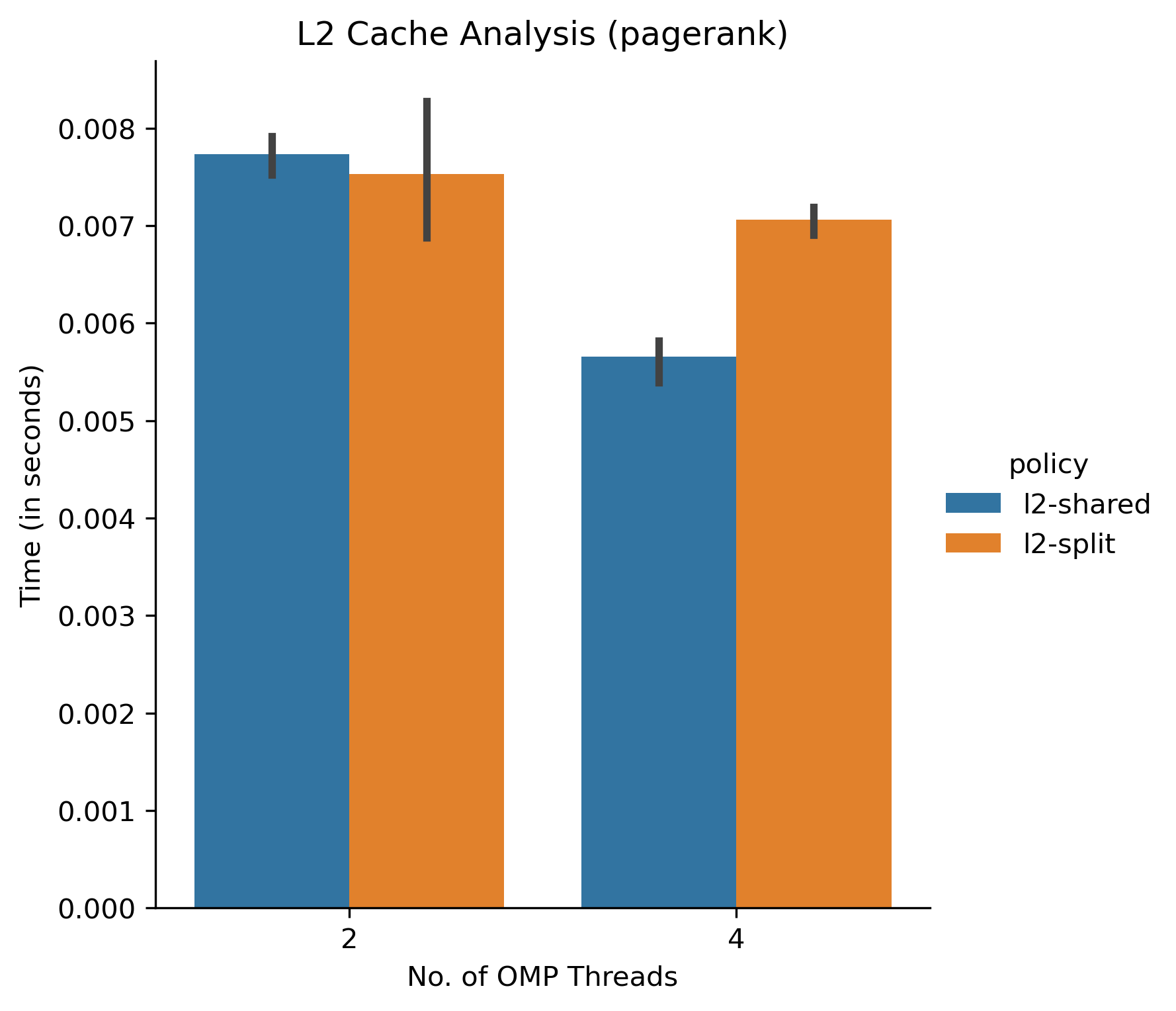}
         \caption{Page Rank}
         \label{fig:crono-pr}
     \end{subfigure}
      \hfill
     \begin{subfigure}[b]{0.24\textwidth}
         \centering
         \includegraphics[width=\textwidth]{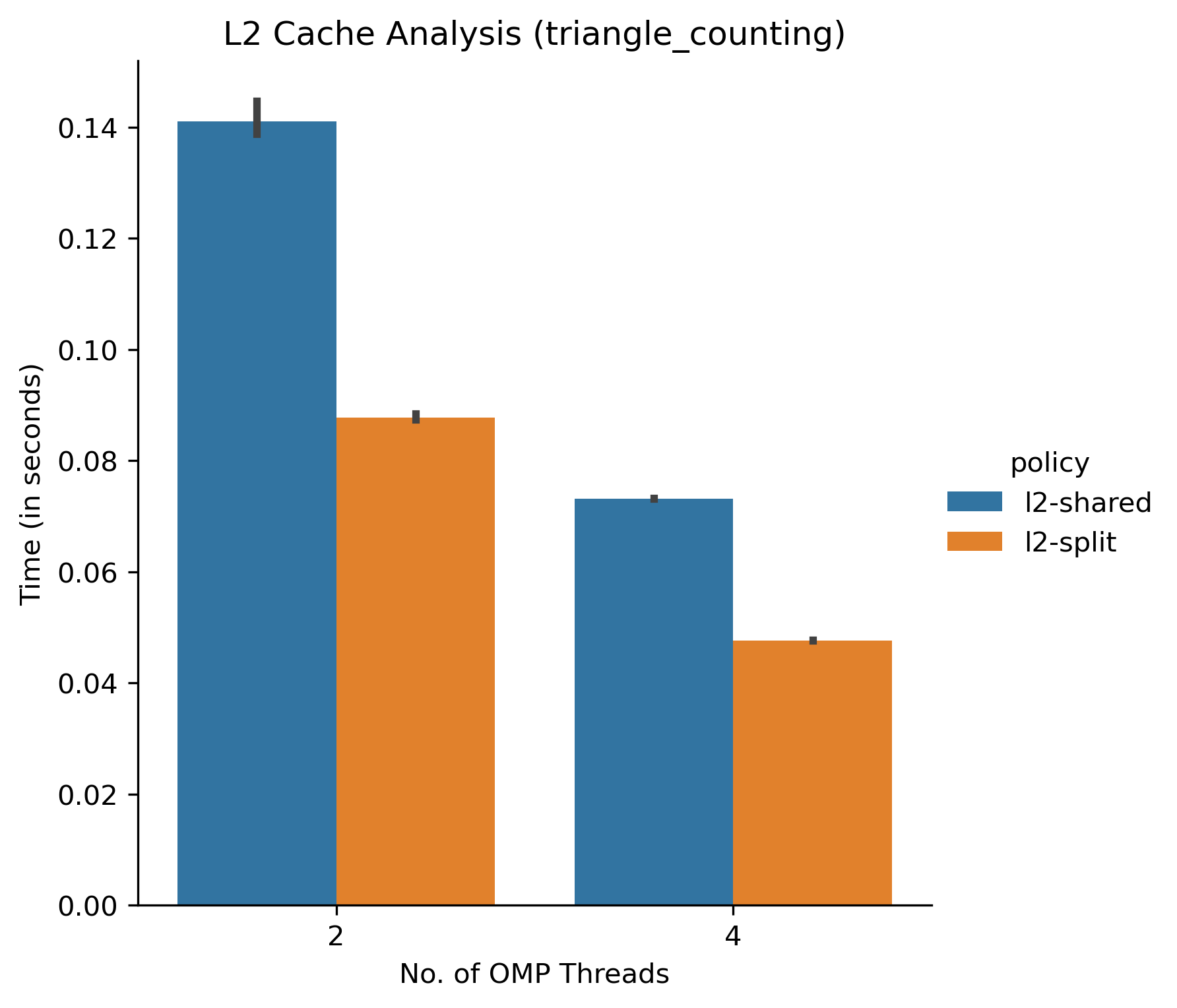}
         \caption{Triangle Count}
         \label{fig:crono-tc}
     \end{subfigure}
        \caption{L2 impact on CRONO Benchmarks showing differences between \textit{l2-split} and \textit{l2-shared} thread placements}
        \label{fig:l2-crono}
\end{figure}

Finally, we have evaluated some lock based parallel workloads from CRONO benchmark. In Fig \ref{fig:l2-crono}, we can see BFS (Fig \ref{fig:crono-bfs}) and Triangle Count (Fig \ref{fig:crono-tc}) show that these workloads have better \textit{l2-split} performance than \textit{l2-shared}. However, Connected Components (Fig \ref{fig:crono-cc}) and Page Rank (Fig \ref{fig:crono-pr}) show better \textit{l2-shared} performance than \textit{l2-split} but only in the case of 4 threads.

\subsection{Unaffected Workloads}

\begin{figure}[h]
     \centering
     \begin{subfigure}[b]{0.24\textwidth}
         \centering
         \includegraphics[width=\textwidth]{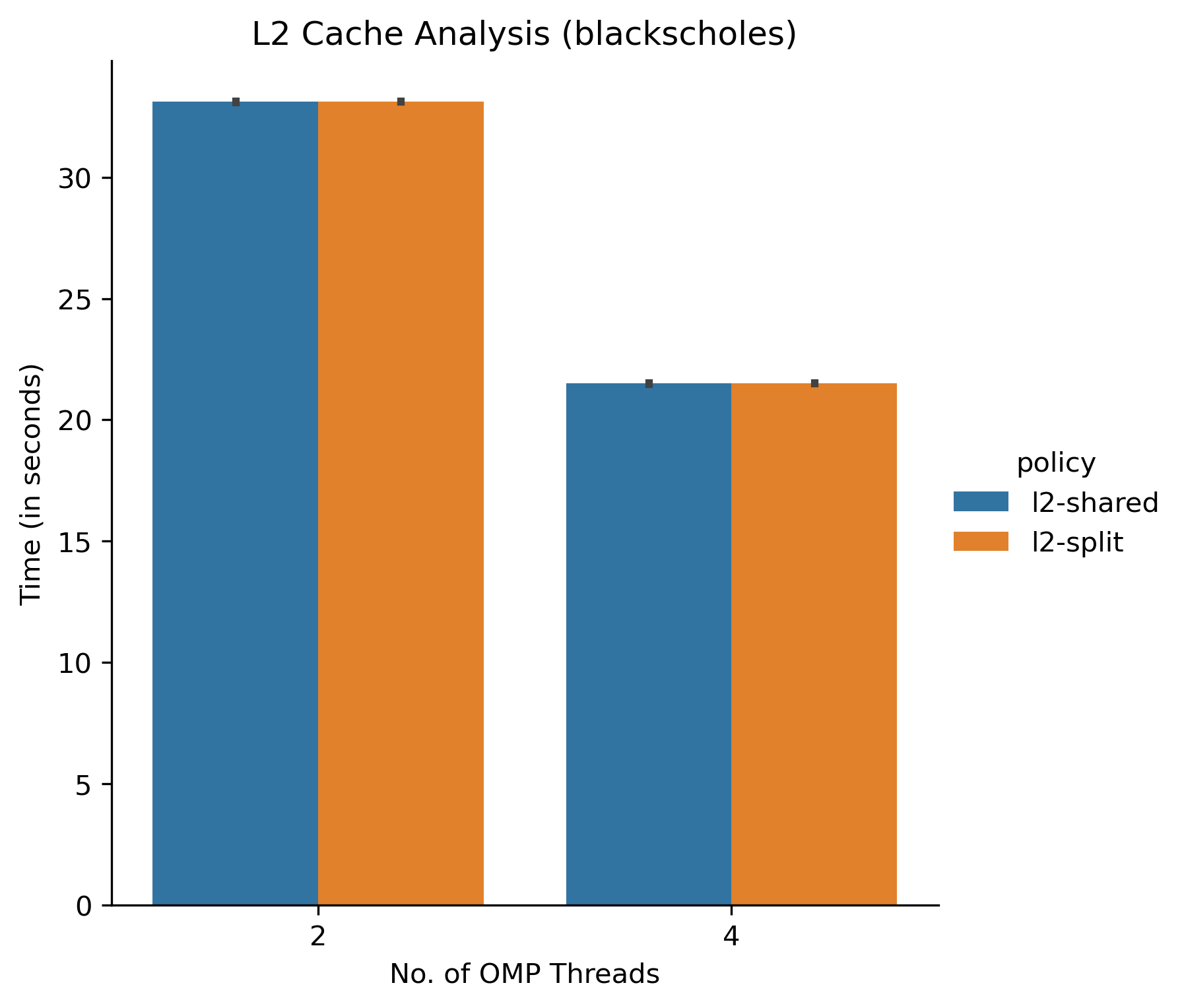}
         \caption{blackscholes}
         \label{fig:parsec-bs}
     \end{subfigure}
     \hfill
     \begin{subfigure}[b]{0.24\textwidth}
         \centering
         \includegraphics[width=\textwidth]{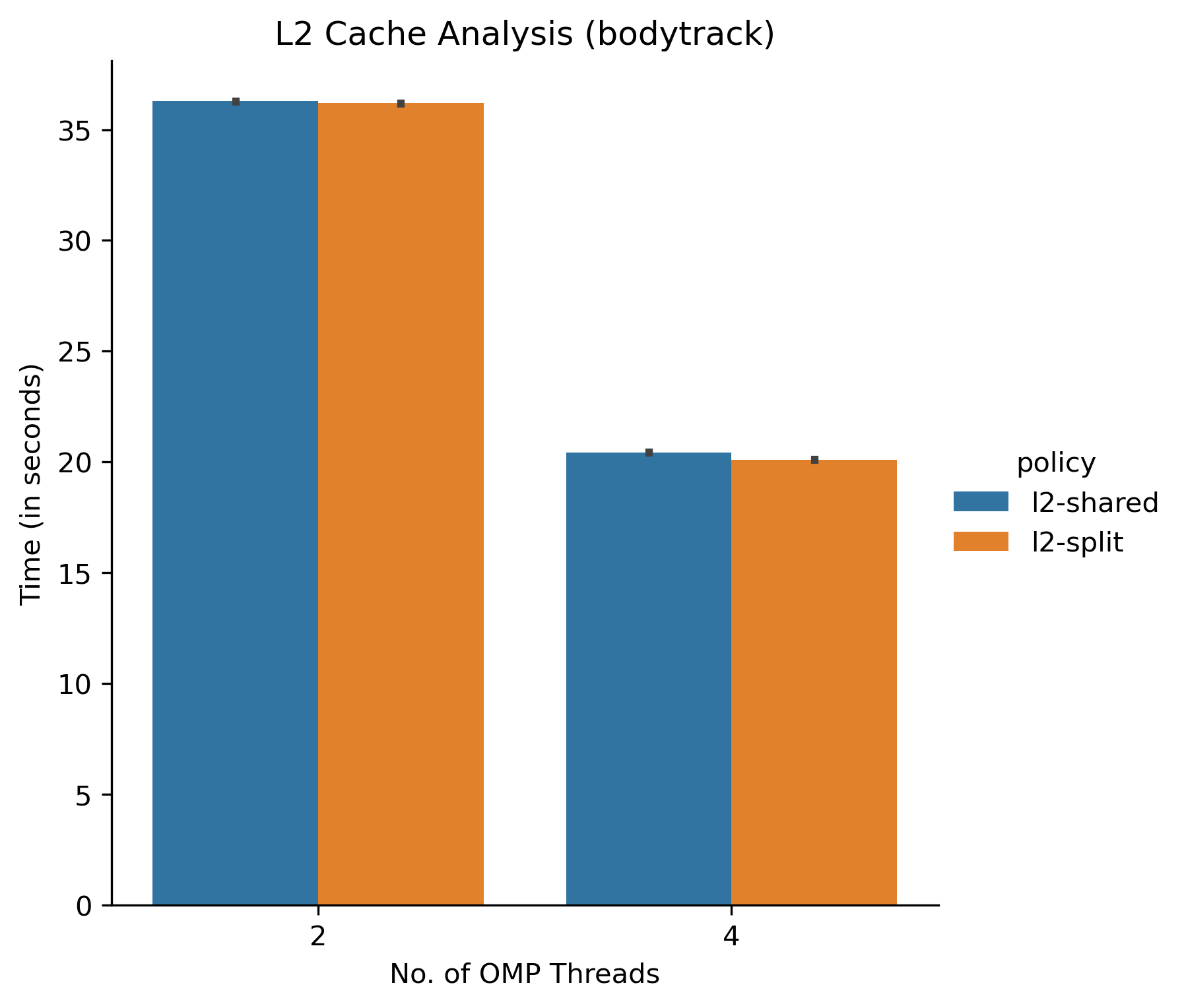}
         \caption{bodytrack}
         \label{fig:parsec-bt}
     \end{subfigure}
      \hfill
     \begin{subfigure}[b]{0.24\textwidth}
         \centering
         \includegraphics[width=\textwidth]{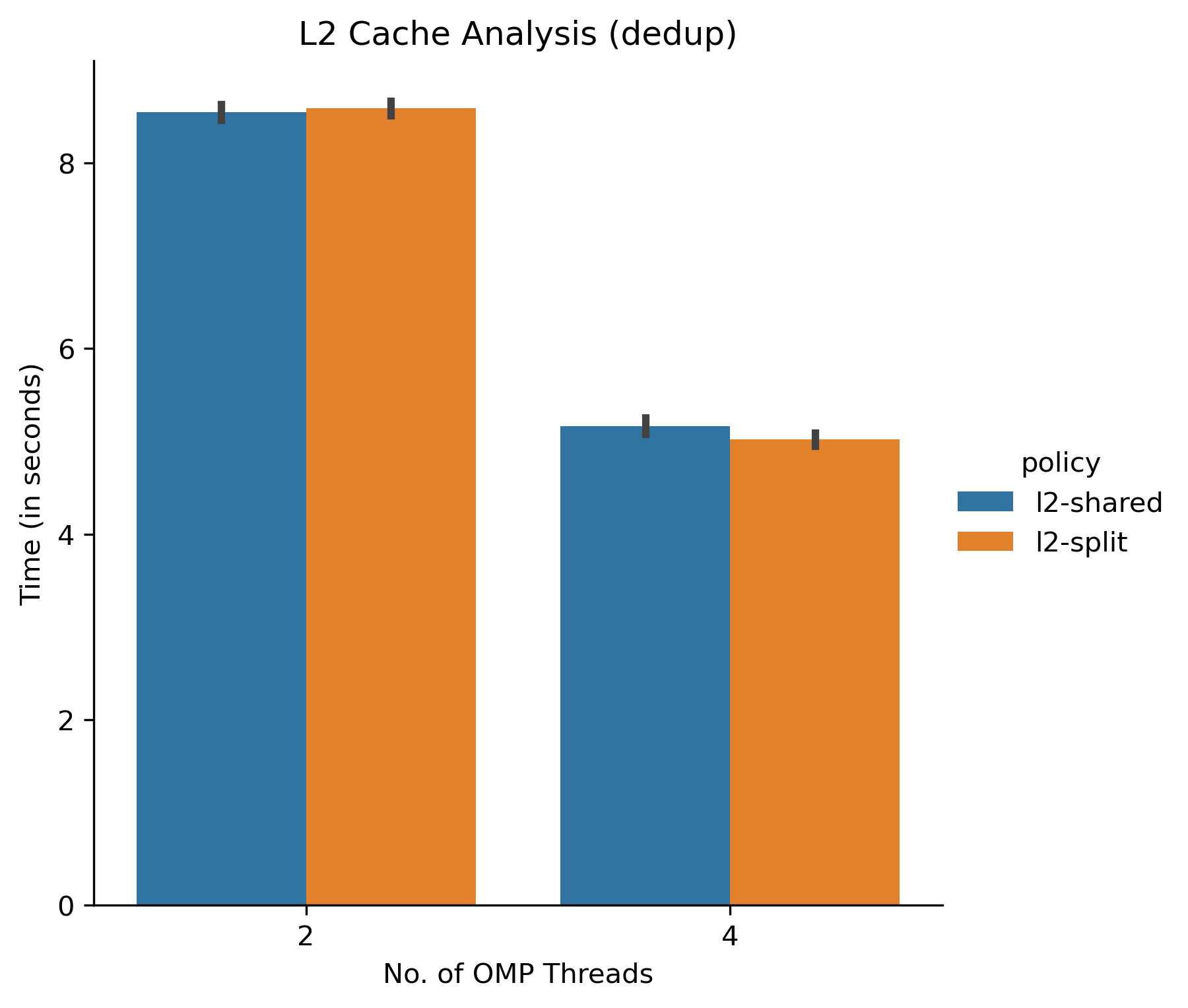}
         \caption{dedup}
         \label{fig:parsec-dd}
     \end{subfigure}
      \hfill
     \begin{subfigure}[b]{0.24\textwidth}
         \centering
         \includegraphics[width=\textwidth]{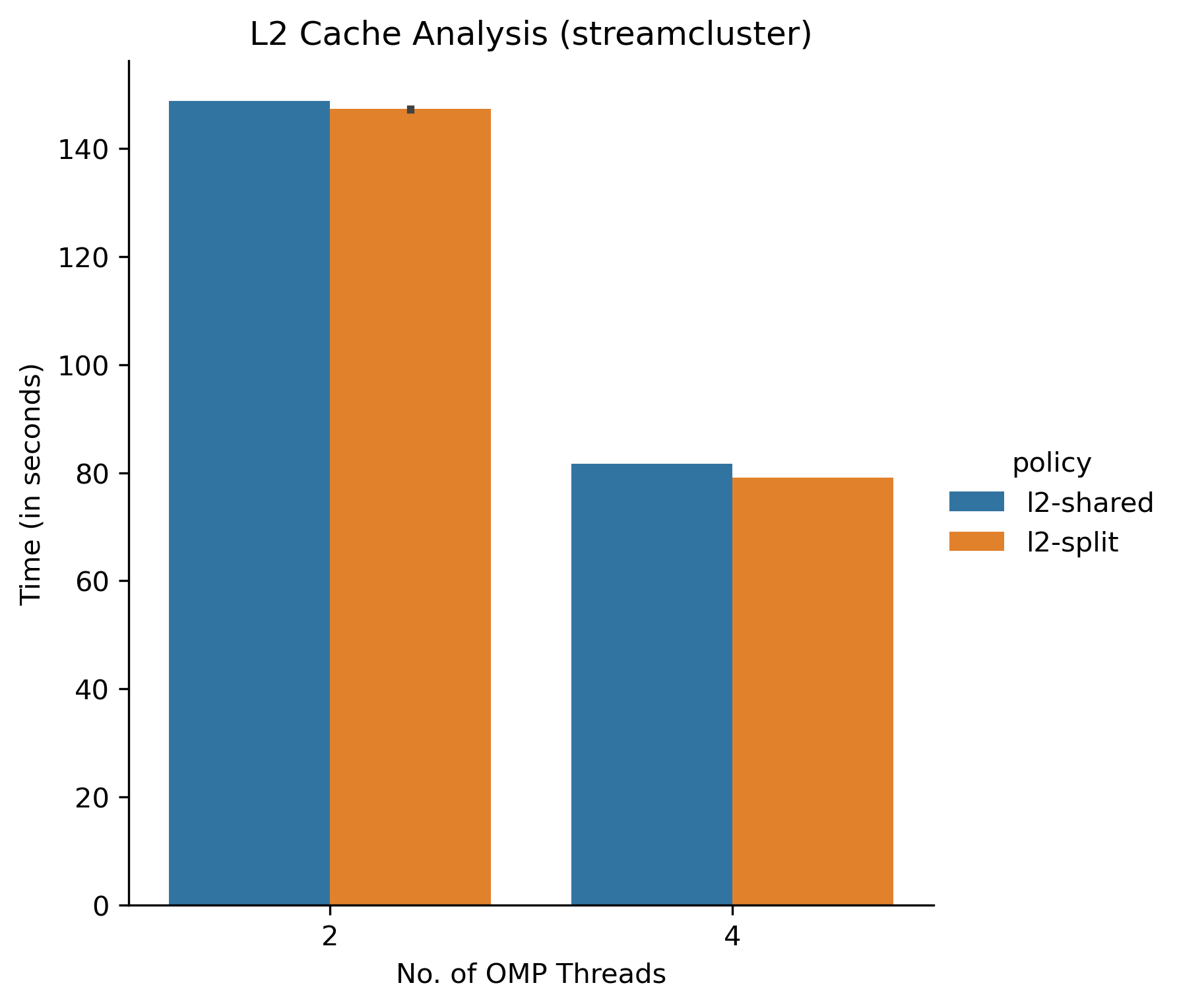}
         \caption{streamcluster}
         \label{fig:parsec-sc}
     \end{subfigure}
        \caption{L2 impact on PARSEC Benchmarks showing no difference between \textit{l2-split} and \textit{l2-shared} thread placements}
        \label{fig:l2-parsec}
\end{figure}

PARSEC benchmark \textit{oddly} shows no difference when compared between \textit{l2-split} and \textit{l2-shared} and even with the scaling up of no. of threads which is captured in Fig \ref{fig:l2-parsec}. This only implies whatever shared cache impact these workloads have, it seems to be almost negligible, if not none.

\section{Explanation of the Impact of Hybrid Cache}

In this aspect, hybrid cache seems to have only minimal impact on data-shared parallel workloads. Moreover, as discussed in the previous section, some workloads (SPMM from HiParTi and CC,PR from CRONO) do exhibit some improvement when threads are sharing cache versus not. However, this difference in performance so small that we cannot make any definite conclusions about our hypothesis. On contrast, whatever difference we observe, it may be due to the cost of coherency protocol which is not relevant to our hypothesis.

\section{Related Work}

Hybrid CPU architecture have been studied before in terms of scientific computing \cite{bib-rel1}, \cite{bib-rel2}, databases \cite{bib-rel3}, etc. They are even explored in domains such as programming models \cite{bib-rel4}, \cite{bib-rel8}, compilers \cite{bib-rel5}, execution frameworks \cite{bib-rel6}, \cite{bib-rel7}, etc. However, most if not all of these works focus on either hybrid CPU-GPU or CPU-FPGA systems.

The idea of big.LITTLE type of heterogeneous core architecture as been explored almost more than a decade ago and change the mobile computing landscape \cite{bib-rel9}. This resulted in an decade worth of research explorations of hybrid core architecture in simulators \cite{bib-rel10}, task allocation/scheduling \cite{bib-rel11}, \cite{bib-rel16}, power modeling \cite{bib-rel12}, load balancing \cite{bib-rel13}, AI workloads \cite{bib-rel14}, \cite{bib-rel15}, etc. But, even such efforts didn't consider exploring the hybrid core architecture in HPC.

There are relevant research efforts that are close to HPC such as high-performance many-core on-chip systems \cite{bib-rel17}, HPC/Cloud infrastructure \cite{bib-rel18}, power/energy \cite{bib-rel19}, etc. However, none of these studies focus on impact hybrid cores in terms of execution, time, memory and power on real-world HPC applications.

\section{Future Work}

This work is part of a larger research effort on studying Hybrid CPU and memory architectures and their impacts on performance of parallel workloads in terms of execution time, memory usage and power consumption. In this paper we have only mentioned the execution time aspect of Hybrid CPU impact. However, we have an on-going effort, building on top of this work, to conduct the same type of study using the same methodologies on Cloud/Server applications running on a Hybrid Server CPU architecture to explore the potential of Hybrid Server architectures in Cloud.

We also have an on-going research effort to analyze memory access traces and power event traces to study the relationship between memory and power. The main goal of this effort is to model the memory access patterns of a parallel workload along with the power information and predict thread placements that are power efficient over time.

\section{Conclusion}

In this paper we have, quantitatively and qualitatively, studied the impact of hybrid CPU architecture both in terms of 1) impact of hybrid cores and 2) impact of hybrid cache on parallel HPC workloads. We show that, the parallel applications with work imbalance scale better with hybrid cores, when thread affinity is turned off and scheduler is allowed to dynamically assign threads to cores. We also have explored the potential impact of hybrid cache architecture, however, there seems to be very minimal impact and even that may be due to cost of coherence protocol when a data-shared threads are co-located vs not co-located. We think these are novel observations and highlight the potential of using hybrid CPUs for efficiently running parallel HPC and Cloud applications.

\end{document}